\documentclass[preprint]{aastex}
\begin{document}

\title{The Local Tully-Fisher Relation for Dwarf Galaxies}

\author{Igor D. Karachentsev,
Elena I. Kaisina,
Olga G. Kashibadze (Nasonova)}
\affil{Special Astrophysical Observatory, Nizhniy Arkhyz, Karachai-Cherkessia
369167, Russia}

\begin{abstract}
We study different incarnations of the Tully-Fisher (TF) relation for the Local
Volume (LV) galaxies taken from Updated Nearby Galaxy Catalog. The UNGC sample
contains 656 galaxies with $W_{50}$ HI-line-width estimates, mostly belonging to
low mass dwarfs. Of them, 296 objects have distances measured with accuracy
better than 10\%. For the sample of 331 LV galaxies having baryonic masses $\log
M_{bar} > 5.8 \log M_\sun$ we obtain a relation $\log M_{bar}= 2.49 \log W_{50}
+ 3.97$ with observed scatter of 0.38 dex. The largest factors affecting the
scatter are observational errors in $K$-band magnitudes and $W_{50}$ line widths
for the tiny dwarfs, as well as uncertainty of their inclinations. We find that
accounting for the surface brightness of the LV galaxies, or their gas fraction, or
specific star formation rate, or the isolation index do not reduce essentially 
the observed scatter on the baryonic TF-diagram. We also notice that a sample of 
71 dSph satellites of the Milky Way and M31 with known stellar velocity dispersion 
$\sigma^*$ tends to follow nearly the same bTF relation, having slightly lower 
masses than that of late-type dwarfs.
\end{abstract}


\section{Introduction}
Since the first wholesale observations of galaxies in the HI 21 cm line, some
tight correlations between different integral properties of spiral galaxies were
exposed (Roberts 1969, Balkowski et al. 1974). Later Tully \& Fisher (1977)
noticed the clear power-law correlation between HI 21 cm line width and optical
luminosity of galaxies, and suggested to use this dependence to measure
distances of galaxies. Since then, more than a thousand articles have appeared,
addressing to the Tully-Fisher (=TF) relation in different bands of optical and
infrared wavelength with different indicators of inner rotation amplitudes.
Tully \& Pierce (2000) have calibrated the TF relation in $B$, $R$, $I$, and $K$
bands, mentioning that the slope of the TF relation grows systematically with
wavelength. According to these authors, scatter on the diagram is small for
massive disk galaxies, allowing to measure distances of spiral galaxies with an
accuracy of about 20\%. A linear relation between HI line width $W_{50}$
measured at the 50\% level of the maximum and standard optical diameter for flat
edge-on galaxies (Karachentsev 1989; Karachentsev et al. 1999) is a particular
instance of the TF relation. For Sc--Sd galaxies with apparent axis ratio
$a/b>7$, this method also yields the distance accuracy of $\sim 20$\%.

Later, McGaugh and coauthors noted that TF relation connecting mass of a dark
halo and baryonic mass of a galaxy should additionally account for mass of a
gas component (McGaugh 2005; 2012, McGaugh et al. 2010). This correction turned
out to be mostly significant for dwarf irregular galaxies evolving slower than
spirals, with extant primordial gas presumably prevailing over star component.
This version of Tully-Fisher diagram is known as the baryonic TF (= bTF)
relation.

The observed scatter in the TF diagram keeps steadily growing from massive
spirals to dwarf galaxies. The main reason of the scatter is the growth of
observational uncertainties in apparent magnitudes and line widths $\log W_{50}$
increasing towards dwarfs. Another reason may be the shallow potential well of a
dwarf galaxy where gas can easily escape due to some external factors. One more
reason for the scatter may be the irregular shape of dwarfs making it difficult
to determine an inclination of rotation axis to the line of sight. Moreover, HI
imaging study of low-mass dIr galaxies performed with the Giant Metrewave Radio
Telescope (GMRT) showed that a direction of HI major axis for some tiny dwarfs
does not always coincide with the direction of their optical major axis (Begum
et al. 2008a). It should be noted that the neutral hydrogen layers of many
galaxies may be warped: the inclination may significantly change in the outer
parts (e.g., Garcia-Ruiz \& Sancisi 2002). Optical inclinations are often
inappropriate for warped HI discs. These circumstances make rather uncertain the
correction of rotation amplitude for dwarf galaxy inclination unless spatially
resolved HI observations are available. Finally, dwarf
galaxies usually have slowly rising rotation curves and reach the $V_{flat}$
value in hardly detected peripheric regions. As a result, their
$V_m=W_{50}/(2\times \sin i)$ corresponds to only a part of the full rotation
amplitude (Swaters et al. 2009). It should be also noted that bursts of star
formation in a dwarf galaxy affect its integral luminosity more significantly
than in a massive galaxy, causing one more source of scatter towards dwarf
sector of the TF diagram.

The main goal of our paper is to improve a calibration of the TF relation on its
low-mass end, using the local dwarf galaxies with accurately measured distances,
that allows to up-date the distances of many remaining nearby dwarfs with
already known HI line widths.

\section{The Local Volume sample.}
For obvious reasons, the most representative and homogeneous sample of dwarf
galaxies can be obtained examining the closest volume of the Universe.
Kraan-Korteweg \& Tammann (1979) have made the first attempt to compile such a
sample. Their list included 179 galaxies with radial velocities
$V_{LG}<500$~km/s relative to the centroid of the Local Group. Due to the
systematic sky surveys in the optical range and HI 21 cm line, the number of
nearby galaxies with similar velocities grew rapidly. 25 years later, the
Catalog of Neighboring Galaxies (=CNG, Karachentsev et al. 2004) amounted to 450
galaxies in the volume limited to depth of 10 Mpc. The refined and up-to-date
version of this catalog, Updated Nearby Galaxy Catalog (=UNGC, Karachentsev et
al. 2013) contains 869 galaxies with individual distance estimates $D<11$~Mpc or
radial velocities $V_{LG}<600$~km/s. More than 85\% of this sample are dwarf
galaxies with luminosities less than those of Magellanic Clouds. Various
observational data on these objects including their images are compiled in the
Catalog \& Atlas of the Local Volume Galaxies (Kaisina et al. 2012,
http://www.sao.ru/lv/lvgdb/). The database is still being enlarged with new,
mainly dwarf, galaxies and contains (as on September 2016) already 1049 objects
distributed all over the whole sky. We use this sample below to construct the TF
diagram in its different incarnations.

The data on HI fluxes of galaxies, $F(HI)$, expressed in HI-magnitudes
$m_{21}=17.4-2.5\log F(HI)$ according to Paturel et al. (1997), as well as HI 21
cm line widths, $W_{50}$, originate mainly from HI sky surveys: the HI Parkes
All Sky Survey (=HIPASS; Koribalsky et al. 2004; Meyer et al. 2004; Wong et al.
2006; Staveley-Smith et al. 2016), Arecibo Legacy Fast ALFA Survey (=ALFALFA;
Giovanelli et al. 2005; Haynes et al. 2011), Westerbork Survey (Kova\^{c} et al.
2009) and special observations of selected dwarf galaxies (Huchtmeier et al.
2000, 2001, 2003; Begum et al. 2008a). The individual references to these and
other HI data are represented in the Local Volume (=LV) database
(http://www.sao.ru/lv/lvgdb/).

Table 1 presents numbers of the LV galaxies with distances measured via
different methods. The first column corresponds to the total UNGC sample updated
with recent observations, the second column refers to UNGC objects detected in
the HI line. The first row shows the number of galaxies whose distances are
measured with errors less than 10\% : from the tip of red giant branch (TRGB),
Cepheid luminosity (Cep), Supernovae (SN), RR Lyra variables (RR),
color-magnitude diagram (CMD), and the horizontal branch (HB). The second row
gives the number of galaxies with distance estimates obtained from surface
brightness fluctuations (SBF). The accuracy of this method, according to Tonry
et al. (2001), is also about 10\%, but this method is suitable for early-type
galaxies with low amount of dust and gas, leading to the risk that the $W_{50}$
value for them can differ systematically from that of late-type galaxies of the
same luminosity. The third row of the table refers to galaxies with distances
determined from their membership (mem) in groups, where other (usually brighter)
members have individual distance estimates. According to recent TRGB-distance
measurements for galaxies with earlier fixed membership, the mean distance error
for this category of galaxies is estimated by us as 17\%. Galaxies with
distances determined from the ordinary Tully-Fisher relation (TF) or its
baryonic version (bTF) are denoted in the fourth row. The fifth row corresponds
to galaxies with distance estimates collected from less reliable methods:
luminosity of the brightest stars (bs), planetary nebula luminosity function
(PN) or from apparent texture of an object (txt). The sixth row shows the number
of galaxies with kinematic distance estimates obtained from their radial
velocities where the Hubble parameter is set to $H_0=73$~km~s$^{-1}$~Mpc$^{-1}$
and the virgocentric velocities are factored (h') or not (h). Finally, the total
number of the LV galaxies with different distance estimates is shown in the last
row.

\begin{table}
\caption{Numbers of the Local Volume galaxies with distances measured by
different methods.}
\begin{tabular}{lcc}
\hline
Method & $ N_{all}$ & $N_{HI}$\\
\hline
TRGB, Cep, SN, RR, CMD, HB & 459 & 296\\
SBF & 18 & 9\\
mem & 269 & 85\\
TF, bTF & 189 & 188\\
bs, PN, txt & 38 & 24\\
h, h' & 76 & 54\\
\hline
All methods & 1049 & 656 \\
\hline
\end{tabular}
\end{table}

As seen, our LV sample contains a total of 656 galaxies with estimates of
$W_{50}$ and $m_{21}$. Almost half of them have distances measured with an
accuracy of $5-10$\%. This circumstance is particularly significant for nearby
galaxies, because their small radial velocities can't be a robust indicator of
distance being affected by peculiar motions.

A literature review shows that authors often preselect galaxies by some specific
properties complicating the interpretation of TF diagram. For instance,
Papastergis et al. (2016) in their study of the baryonic TF relation for heavily
gas-dominated ALFALFA objects, selected only galaxies with apparent axial ratio
less than 0.25, omitting a huge number of low mass galaxies which are gas-rich
but intrinsically thick. Sometimes authors do not completely specify their
selection criteria or neglect galaxies deviating significantly from the TF
regression line. It results in artificially reducing scatter in the TF relation.
Below, we look toward keeping in our sample as many objects as possible,
eliminating only face-on disks, or galaxies located in the Zone of Avoidance, or
those with ambiguous distance estimates. In particular, we did not omit from our
analisys a case of nearby bright galaxy M82 having accurate photometric data,
the reliable estimate of the distance, $W_{50}$ width and inclination $i$,
albeit this is a known starbust galaxy involved in a strong interaction with M81
impressively visible in HI (Yun et al. 1994).

\section{Intrinsic axial ratio and inclination corrections.}
The distribution of apparent axis ratios for the LV galaxies of different
morphological types is shown in Figure~1. The galaxies are marked by small
circles. Many of them overlap with each other. The large circles and vertical
rectangles indicate the mean values and the standard deviations for each type in
de Vaucouleurs scale. The left side of the diagram ( $T < 1$) contains mostly HI
undetected objects. In general, the intrinsic axial ratio is larger for
lenticular and irregular galaxies than for spirals, and among the spirals is
smaller for the late types than for the early ones. The minimum flattening
corresponds to the Sd galaxies (T = 7), although a thin edge-on galaxy would 
be classified as Sd instead of Sc or Sm since one cannot appreciate the degree
of structure in the pattern at $i \simeq 90$ deg.

As it was already mentioned, uncertainties in correction for inclination of
dwarf galaxies may cause scatter in the TF diagram. For oblate ellipsoids, an
inclination of rotation axis of a galaxy to the line of sight is defined as $$
\sin^2 i =[1-q^2]/[1-q^2_0], \eqno(1)$$ where $q = b/a$ is the apparent axial
ratio. In case of CNG sample, the intrinsic flattenings $q_0 = (b/a)_0$ were
taken equal to 0.07 for morphological types $T$ = 6 or 7, 0.12 for types 5 or 8,
0.18 for types 4 or 9 and 0.20 for all other morphological types, according to
the Second Reference Catalog (RC2) by de Vaucouleurs et al. (1976). This
relation between the intrinsic axial ratio and de Vaucouleurs morphological type
is represented in Fig.~1 by a dotted line with open circles. Analyzing the apparent axial
ratio statistics for dwarf galaxies taken from the UNGC, Roychowdhury et al.
(2013) confirmed the known fact that disks of dwarf galaxies are much thicker
than it was assumed in RC2. These authors obtained the mean intrinsic flattening
of 0.57. Just the same value was found by Sanchez-Janssen et al. (2016) for
dwarf population of the Virgo cluster. The authors of both papers noted that the
apparent axial ratio statistics for dwarf galaxies is better described by a
model of oblate triaxial spheroid with axes ratio 1.00 : 0.94 : 0.57. Also,
Roychowdhury et al. (2010) derived approximately the same axial ratio for HI
images of dIr galaxies based on FIGGS data from GMRT (Begum et al. 2008a).

Paturel et al. (1997) suggested a refined set of $q_0$ values depending on
morphological type $T$, which was used in UNGC to determine the inclination $i$
from relation (1): $$ \log(q_0) = -0.43-0.053\times T\,\,\,\,\ {\rm for}\,\,\
T\leq 8 \eqno(2) $$ $$ \log(q_0)= -0.38 \,\,\,\,\ {\rm for}\,\, T = 9,10.$$ The
behaviour of intrinsic flattenings versus morphological type in this case is
depicted by the dashed line with grey circles in the Fig.~1. The recipe of 
Paturel et al. (1997) for $q_0$ seems to be significantly more realistic than 
the previous scheme used in RC2. Yet, the condition (2) presumably underestimates 
the actual thickness of the disks of dwarf galaxies, resulting in their inclination 
$i$ estimates. Here we should point out a big leap in the condition (2) passing 
from morphological type $T$ = 8 to 9. A typical scatter in morphological classification 
of $\Delta T=\pm1$ can lead to a significant error in galaxy inclination.

Yuan \& Zhu (2004) analyzed a sample of 14988 disk galaxies taken from LEDA
database and derived for them the relation $$ \log(q_0) = -0.580 - 0.067\times
T\,\,\,\,\ {\rm for}\,\,\ T\leq 7 \eqno(3) $$ $$ \log(q_0)= -2.309 + 0.185\times
T\ \,\,\,\,\ {\rm for}\,\, T = 8,9,10.$$ This distribution is shown in the
Figure~1 by the solid line with black circles. It looks as intermedian between 
those adopted in CNG and UNGC samples for most morphological types.

There is another way to evaluate the intrinsic axial ratio using stellar mass
instead of optical morphology. Sanchez-Janssen et al. (2010) investigated the
role of stellar mass in shaping the intrinsic thickness of dwarf systems. They
found that the intrinsic axial ratio varies with stellar mass in a parabolic
fashion having the minimum value near the stellar mass $M^* = 2\times 10^9
M_\sun$. In our LV sample the distribution of all 1049 galaxies according to
their total luminosity in K-band, $L_K$, and morphological type is shown in
Figure~2. Assuming an approximate ratio $M^*/L_K \simeq 1 M_\sun/L_\sun$ (Bell
et al. 2003) this diagram can be transformed into the $M^*$ vs. T distribution.
The mean values of $L_K$ and standard deviations are marked, like in Fig.~1, by
large circles and rectangles. The shape of diagram resembles parabola with the
maximum near Sb-type and a scatter increasing towards both the sides.

The number distribution of all LV galaxies with their K-band luminosity is
illustrated by Figure~3. Among them, the 656 HI-detected galaxies are marked by
dark color. As one can see, a fraction of the LV galaxies with the HI detections
decreases systematically from bright to faint objects. More than 99\% of the HI
detected galaxies are concentrated within the luminosity range of $\log(L_K)$ =
6.0 - 11.0. Such a segregation along $L_K$- axis may be a source of specific
bias in the statistics of apparent and intrinsic flattenings.

Figure~4 represents the distribution of 1049 LV galaxies with their apparent
axial ratio and K-band luminosity. Galaxies detected and undetected in HI are
indicated by filled and open small circles, respectively. The large circles and
vertical rectangles correspond to the average value of $q$ and its dispersion in
bins of $\Delta \log(L_K)$ = 1.0. The diagram shows a clear tendency for minimal
values of $q$ to follow the parabolic envelope line with the absolute minimum of
$q = 0.05$ near $\log(L_K) \simeq 9$. Some dwarf gasless galaxies, like UMaI,
deviate from the envelope line. They usually locate close to a massive
neighboring galaxy (Martin et al. 2008), and their alongated shape can be caused
by tidal perturbation.

Bradford et al. (2016) explored a sample of 930 isolated galaxies with axial
ratios determined from the Sloan Digital Sky Survey (SDSS) and suggested for
them a relation between true flattening and stellar mass of a galaxy as: $q_0 =
0.85-0.057\times \log(M^*/M_\sun)$. According to this formula, the $q_0$ value
grows monotonically from 0.2 for massive galaxies up to 0.5 for dwarfs. However,
there is no place still for the common fact that the minimal values $q_0 \simeq
0.05$ correspond just to the low mass galaxies of Sd-Sdm types.

Following Sanchez-Janssen et al. (2010), Lelli et al. (2016a) adopted the
parabolic relation: $$ q_0 = 5.2125 - 1.125\log(M^*) +
0.0625\log(M^*)^2,\eqno(4) $$ which reaches a minimum value of $q_0 = 0.15$ at
$\log(M^*) = 9.0$ and gives $q_0 = 0.4$ at $\log(M^*)$ equals 7.0 and 11.0. Both
the relations by Bradford et al.(2016) and Lelli et al. (2016a) are shown in
Figure~4 with dotted and dashed lines respectively.

Exploring properties of the LV galaxies and taking into account axial ratio
statistics for ultra-flat galaxies (Karachentseva et al. 2016), we
re-parametrized the parabolic relation and found the optimum parameters: $$q_0
= 5.128 - 1.114\log(M^*) + 0.0612\log(M^*)^2.\eqno(5) $$ This modified parabolic
distribution, shown in Fig.~4 with solid line, reaches a minimum value of $q_0 =
0.059$ at $\log(M^*) = 9.10$. We have to stress here that equations (4) and (5)
should not be extrapolated for stellar masses below $\sim 10^6 M_{sun}$ since
they give unphysical results ($q_0 > 1$).

If the distribution of galaxies over their intrinsic flatennings as a function
of type or stellar mass is specified correctly, than, in the case of random
spatial orientation of galaxy planes, the distribution of inclination angles
will correspond to the $\sin i$ -law (Yuan \& Zhu, 2004). Fig.~5 represents the
distribution of Local Volume galaxies over inclination $i$ with step of
$10^{\circ}$ under different assumptions (2)-(5) about intrinsic axial ratios.
The solid line traces the expected distribution with uniformly random
orientation of galaxy spin vectors over the sky. Vertical bars denote statistical
errors.

As one can see from Fig.~5, the model (2) for galaxy distribution over intrinsic
axial ratios as well as cases (4) and (5), reproduces the expected $\sin i$ -law
much better than the RC2 scheme adopted in the CNG. The most significant
deviations from the expected law in all the cases owe to underestimating the
number of face-on galaxies: $i<20^{\circ}$. However, we have to exclude such
galaxies from further consideration due to large uncertainties in corrected HI
line width: $W^c_{50}= W_{50}/\sin i$. Another significant differences are seen
in $i>70^{\circ}$ sector, but it has little effect on the corrected HI line
width. One may suggest that the observed mismatch with the $\sin i$-law may be
driven by E, S0, and dSph galaxies. To test this hypothesis, we considered
in the UNGC sample only galaxies with $ M^*> 10^6 M_{sun}$ and $T > 0$. As seen,
this case shown by grey squares does not reduce the mismatch noticeably.
Finally, the modified parabolic relation (5) connecting $q_0$ with $M^*$
is used by us in further study.

\section{Stellar Tully-Fisher relation for dwarfs.}
Considering different kinds of Tully-Fisher relation for the Local Volume,
dominated by objects with low mass and luminosity, we have confined ourselves to
the sample of galaxies with accurate distances (TRGB, Cep, SN, RR, CMD, HB).
Then, we excluded from this sample two galaxies: Maffei2 and HIZSS03 with
Galactic extinction of $A^G_B > 3.0^m$ according to Schlafly \& Finkbeiner
(2011). To reduce errors caused by the uncertainty in $W_{50}$ correction due to
a galaxy inclination $i$, we further consider only galaxies with $ i > i_{lim} =
45^{\circ}$. This condition decreases our sample from 296 to 206.
The analysis shows that stronger limitation does not change significantly the
scatter in the TF diagram, but notably diminishes the sample volume.

Fig.~6 reproduces the classical Tully-Fisher diagram as the relation between the
blue absolute magnitude of a galaxy, corrected for Galactic and internal
extinction, and the HI line width for 206 galaxies taken from the LV database.
The internal extinction was accounted according to Verheijen \& Sancisi (2001):

$$ A^i_B=[1.57+2.75(\log W_{50} - 2.5)]\log(a/b), \eqno(6)$$ if $W_{50} > 78$ km
s$^{-1}$, otherwise $ A^i_B=0$.

In the upper panel of Fig.~6, the measured width $W_{50}$ is plotted as abscissa
covering the range from 10 to 500 km/s. The median value of $W_{50}$ for this
sample amounts to 54 km/s which is much lower than that for other TF-samples
examined by Geha et al. 2006; Bradford et al. 2015, 2016; Brook et al. 2016;
Sales et al. 2016; Papastergis et al. 2016. The $M_B$ vs $\log(W_{50})$ relation
is well described by the linear regression (a straight line) in the whole range
of $W_{50}$ with a slope of $-6.85$ and a dispersion of $\sigma_M=1.06^m$. Here
and below, the straight line in the TF diagram is a robust fit to the
ensemble with errors in magnitudes. As seen, the dispersion increases
appreciably while passing from massive spirals to dwarfs due mainly to
observational errors. In the bottom panel, the corrected value of width,
$W^c_{50} = W_{50}/\sin i$, is plotted on the horizontal axis. The view of the
diagram has been changed subtly: the slope, $-6.96$, and the scatter, $\sigma_M
=1.10^m$, remained nearly the same.

To evaluate the behaviour of scatter on the TF diagram as a function of limiting
inclination angle $i_{lim}$, we have consistently taken $i_{lim} = 30, 35, ...
80^{\circ}$. For each subsample we determined its new regression line and
calculated dispersion $\sigma_M$ regarding to it. The results are presented in
two panels of Fig.~7. The left panel corresponds to the case when the
inclinations were determined from equation (2) via optical morphology, and the
right panel illustrates the another manner with equations (5) accounting stellar
masses. The numbers over the upper border of both panels indicate galaxy numbers
in each subsample. The solid line and dashed line correspond to observed,
$W_{50}$, and corrected, $W^c_{50}$, line width. These data show that giving
away a half of the initial sample at $i_{lim} = (55 - 60)^{\circ}$, we may
decrease the scatter only by $10-20$\%. Also, the correction of line width for
inclination does not improve noticeably a dispersion on the TF diagram inhabited
mainly by dwarf galaxies. In other words, the $W_{50}$ correction for
inclination in the realm of dwarfs looks as just a formal, optional procedure.
This curious peculiar phenomenon was already noticed by Obreschkow \& Meyer
(2013).

As noted by many authors (Aaronson et al. 1979; Aaronson et al. 1986; Giovanelli
et al. 1997; Tully \& Pierce 2000), the dispersion in the TF diagram decreases
systematically towards long wavelengths where the effects of internal extinction
and star-bursts are not so significant. Near infrared $J$, $H$, $K_s$ photometry
of galaxies fulfilled in the 2MASS all-sky survey (Jarett et al. 2000, 2003)
provides background for constructing TF diagrams for spiral galaxies
(Karachentsev et al. 2002). Yet, due to short expositions, the 2MASS survey has
been found to be undersensitive for detecting low surface brightness objects,
especially those ones having blue stellar population. About a half of the Local
Volume galaxies stayed below the threshold of detectability. For many nearby
dwarf galaxies, the integral $K$ magnitude obtained in 2MASS does not trace the
appreciable contribution of the galaxy's periphery. According to the data on
deep photometry of dwarf galaxies in $K$ and $H$ bands (McCall et al. 2012,
Young et al. 2014), the typical error of these magnitudes for dwarfs in 2MASS is
about $0.6^m$. If the accurate photometry was lacking, $K$ magnitudes in UNGC
were estimated by us from $B$ magnitudes using the correlation between the mean
color index $\langle B-K\rangle$ and morphological type of a galaxy (Jarrett et
al. 2003): $\langle B-K\rangle= 4.10$ for the early morphological types E, S0,
Sa; $\langle B-K\rangle=2.35$ for the late types Sm, Im, BCD, Ir and $\langle
B-K\rangle=4.60-0.25\times T$ for intermediate types $T$ from 3 to 8.

The $K$ band luminosity vs $W_{50}$ or $W^c_{50}$ relation for 206 LV galaxies
having accurately measured distances and $i > 45^{\circ}$ is presented in the
upper and lower panels of Fig.~8, respectively. The galaxies with $K_s$
magnitudes from 2MASS are labeled with filled circles, while galaxies with $K$
estimates made from the mean colour index $\langle B-K\rangle$ and morphological
type $T$ are marked with open ones. These data show once again that galaxies
without 2MASS photometry tend to reside in the left lower part of the diagram.
The median value of $W_{50}$ for them is 36.5~km/s, while for galaxies with 2MASS
magnitude their median is twice as much. Notably, both categories of objects ---
with 2MASS magnitudes and with $K$ estimates from the blue magnitude and
morphological type --- follow the same regression line. Its slope, 2.87, and
dispersion, 0.46 are actually the same in the both panels of Fig.~8.

As it has been found by Noordermeer \& Verheijen (2007) and Schombert (2011)
there is a systematic bias in 2MASS total magnitudes caused by some 2MASS surface 
photometry routines underestimating the luminosity of galaxies. This specific 
bias is especially significant for bright extended galaxies plentiful in the
Local Volume. Since high-mass galaxies are star-dominated, a systematic 
underestimation of their K-luminosity may give shallower TF-slopes.

It should be also mentioned that the scheme of converting $B$ magnitudes to $K$
magnitudes adopted by Jarrett et al. (2003) is quite reliable in most cases.
With a characteristic error of $\Delta T = \pm1$ in morphological typing of a
galaxy, the expected error in $K$ magnitude is $\pm0.25^m$. However, for a
special class of transition dwarf galaxies (dSph/dIr), a morphological typing
mistake: $T=-1$ (dSph) else $T=10$ (dIr) leads to a leap in their $K$ magnitudes
by $1.75^m$. Albeit, there are only five objects of this class in our sample:
ESO~410-005, ESO~294-010, LGS-3, Antlia and HS~117.

\section{Gaseous TF-relation for dwarfs.}
Besides the stellar mass of a galaxy, its hydrogen mass, $$M_{HI} = 2.356\times
10^5\times D^2\times F(HI),$$ also shows a tight correlation with HI line width.
In the relation above the distance $D$ is expressed in Mpc and the flux $F(HI)$
in Jy~km~s$^{-1}$. The distribution of 206 Local Volume galaxies over their
hydrogen mass and HI line width $W_{50}$ is presented in the upper panel of
Fig.~9. The regression line has a slope of 2.08, and the dispersion amounts to
$\sigma(\log M_{HI}) = 0.43$. Passing from $W_{50}$ to $W^c_{50}$ (middle panel)
leaves unchanged the regression slope and the dispersion.

As known, some early-type galaxies and transition type dwarfs strongly deviate
downward from the regression line in the gaseous TF diagram. On the bottom panel
of Fig.~9 we excluded from our sample 16 gas-poor galaxies with colour index
$m_{21}- K > 6.0^m$, which approximately corresponds to the condition
$M_{HI}/M^* < 4$\%. This procedure slightly increases the slope (2.29) and
decreases the dispersion (0.40). So we find that the TF dispersion of hydrogen
mass for the late-type LV galaxies is of the same order as the TF dispersion of
their stellar masses.

Fig.~9 outlines the upper limit for $M_{HI}/(W^c_{50})^2$ value marked as
dashed straight line in the lower panel. According to Zasov (1974), active star
formation in disks of galaxies occurs usually near the threshold for their
gravitational instability. This condition corresponds to the linear relation
between the total mass of gas in a disk and its angular momentum: $M_{HI}\propto
V_m\times A_{25}$, where $A_{25}$ is a standard linear diameter of disk. Taking
into account the scaling relation $A_{25}\propto V_m \propto W_{50}$ for UCNG
galaxies (Karachentsev et al. 2013), the presence of upper limit of
$M_{HI}/(W^c_{50})^2$ value agrees well with the idea of Zasov. Examples of the
galaxies residing close to the dashed line are: NGC6744 ($W_{50}=323$,
$\log(M_{HI})=10.31$, $\log(L_K)=10.91$) and KDG177 ($W_{50}=30$,
$\log(M_{HI})=8.33$, $\log(L_K)=8.22$).

\section{Baryonic TF-relation for the Local Volume galaxies.}
Over the last decade, a consensus has been achieved that a classical
Tully-Fisher relation between the stellar mass (or luminosity) and the rotation
amplitude, definable by dark halo mass, should also be contributed by the gas
mass of a galaxy (McGaugh et al. 2000, Verheijen 2001, McGaugh 2005). Such a
correction is particularly essential for galaxies of extremely low luminosities
where the significant part of the gas has not yet been turned into stars. The TF
relation for the total (stars plus gas) mass is known as baryonic Tully-Fisher
relation (=bTF).

Many authors have explored the slope of bTF relation (Geha et al. 2006; Stark et
al. 2009; Begum et al. 2008b; McGaugh et al. 2010; McGaugh 2012; Gurovich et al.
2010; Reyes et al. 2011; Torres-Flores et al. 2011; Catinella et al. 2012;
Bradford et al. 2015; Lelli et al. 2016b; Papastergis et al. 2016). According to
McGaugh \& Schombert (2015), bTF diagram $\log (M_{bar})\propto \beta \times
\log (V_m)$ has a slope of $\beta = 4.0$ in wide range of galaxy masses, which
is not consistent with the expected value of $\beta \simeq 3.0$ for the standard
cosmological model $\Lambda$CDM. The steep $\beta \simeq 4$ slope has been also
determined by Verheijen 2001, McGaugh 2005, Stark et al. 2009, Catinella et al.
2012, Lelli et al. 2016b, and Papastergis et al. 2016. Yet, other authors have
obtained notably less steep $\beta$ values: 2.2 (Begum et al. 2008b), 2.1
(McCall et al. 2012), 2.5 (Kirby et al. 2012). As it was discussed by Brook et
al. (2016), Brook \& Shankar (2016) and Bradford et al. (2016), the slope value
$\beta$ can vary from 2 up to 4 depending on methodological details of
converting observational values $W_{50}, W_{20}$ or $V_{flat}$ to the amplitude
of rotation curve, $V_{rot}$. Also, as the most gas-rich dwarf galaxies follow
the shallow relation $M_{HI}\propto W^{2.2}_{50}$, than the high or low
proportion of irregular dwarfs in the sample can appreciably affect the slope
$\beta$. Anyway, it must be remembered that the $W_{50}$ line widths derived
from spatially unresolved (single dish) observations give systematically
shallower TF relation slope than flat rotational velocities, $V_{flat}$, from
interferometric HI observations. The reason for this is due to the known effect
that dwarf galaxies have slowly rising rotation curves and reach $V_{flat}$ only
in the outermost regions, where the HI surface densities are low and the S/N
ratio goes down (Swaters et al. 2009; Lelli et al. 2014).

To determine the baryonic mass of a galaxy $$M_{bar}= M^*+M_{gas} =
\Upsilon^*_K\times L_K + \eta\times M_{HI},$$ two parameters should be held
fixed: stellar-mass-to-$K$-luminosity ratio (or a similar ratio for any other
photometric band) and the $\eta$ factor used for converting neutral hydrogen
mass to the total gas mass. The published values of $\Upsilon^* = M^*/L_K$ stay
within the range of $[0.5-1.0]M_\sun/L_\sun$. They are based on different models
of synthetic stellar populations. According to analysis fulfilled by McGaugh \&
Schombert (2014), the optimal value of $\Upsilon^*_K$ is close to $0.6 M_\sun
/L_\sun$, though Just et al. (2015) present the value of $M^*/L_K
=(0.31\pm0.02)M_\sun/L_\sun$ for the solar vicinity from Hipparcos data.

The factor $\eta$ is usually set to 1.33 taking into account the correction for
Helium abundance. According to Fukugita \& Peebles (2004), adding molecular gas
to consideration leads to a value of $\eta=1.85$. However, this higher factor is
typical only for disks of massive galaxies. The data by Young \& Knezek (1989)
and McGaugh \& de Blok (1997) show that molecular-to-atomic hydrogen mass ratio
decreases rapidly towards dwarf galaxies as $$ M_{H2}/M_{HI} =3.7-0.8\times
T+0.043\times T^2,$$ and for dwarfs ($T\geq8$) it does not exceed 0.05.

Yet, the situation with $\eta$ value remains rather uncertain. It is commonly
known (see, for example, Fukugita \& Peebles 2004) that the observable quantity
of baryons in galaxies and in hot intergalactic gas of clusters
($\sim0.005\Omega$) is by an order of magnitude less than expected from the standard
cosmological model of nucleosynthesis ($0.045\Omega$). It is suggested that the
bulk of baryons escaping from observations is distributed as warm or hot
intergalactic gas, which can be partially associated with galaxies. Attempts of
detecting hot gaseous coronas were undertaken based on observations of
absorbtion lines OVII and OVIII in quasar spectra (Stocke et al. 2013).
According to Miller \& Bregman (2015), the mass of ionized gas around the Milky
Way is $(3.8\pm0.3)\times 10^9 M_\sun$ within the radius of 50 kpc, and
$(4.3\pm0.9)\times 10^{10} M_\sun$ within 250 kpc. The last value is matching
with the total stellar mass of the Milky Way $(6\times 10^{10} M_\sun)$ and is
several times more than the mass of neutral hydrogen in our Galaxy. If such warm
gaseous halo is typical for other massive and dwarf galaxies, than the
dimensionless factor $\eta$ can amount to $\sim5$. Varying $\eta$ parameter,
Pfenniger \& Revaz (2005) and Begum et al. (2008b) concluded that the minimum
dispersion in bTF diagram is reached within a wide range of $\eta=(3-10)$. This
result implicitly confirms that a large number of hot baryons may be associated
with galaxies beyond the known amount of neutral gas in them.

The distributions of the Local Volume galaxies over baryonic mass and HI line
width, $W_{50}$ or $W^c_{50}$, are presented in the upper and lower panels of
Fig.~10. For factors converting $M^*$ and $M_{HI}$ to baryonic mass we have
adopted $\Upsilon^*_K = 0.60$ (McGaugh \& Schombert 2014) and $\eta = 1.33$. The
slope of the regression line and the dispersion are $\beta = 2.55, \sigma = 0.36$
(for $W_{50}$) and $\beta=2.61, \sigma=0.38$ (for $W^c_{50}$), respectively.

As we have already mentioned, the correction for inclination in the case of dwarf
galaxies does not play any significant role, so we included into our analysis
another 56 faint dwarf galaxies with $M_B>-16.0^m$ and inclinations
$i<45^{\circ}$ (marked as crosses in Fig.~11). We added also 69 mainly dwarf galaxies with
distance estimates (mem) obtained from their membership in the known groups. The
typical distance error for them is $\sim17$\% or 0.14 dex in the scale of
$M_{bar}$, which is three times less than the observed TF scatter. These objects
with arbitrary inclinations if $M_B>-16.0^m$, or $i>45^{\circ}$ otherwise, are
labeled as open diamonds. The regression line for the complemented
sample of 331 galaxies has a slope of $\beta=2.49$ and dispersion of
$\sigma(\log M_{bar})= 0.38$.

To optimize the choice of $\Upsilon^*_K$ and $\eta$ parameters for the Local
Volume galaxies we calculated their baryonic mass with various values of
$\Upsilon^*_K$ from 0.40 to 1.00 and $\eta$ from 1.33 to 5.0. The RMS deviations
from the regression line for each combination of parameters are presented as
matrix in Table 2. The differences in $\sigma(\Upsilon^*_K, \eta)$ turn out to
be small. The minimum value of dispersion, 0.352, is found along the matrix
diagonal going from $\Upsilon^*_K$= 0.40 and $\eta=2.50$ to $\Upsilon^*_K$=
1.00 and $\eta$= 5.0. This result could be interpreted in favour of the presence
of dark (warm) gas around the LV galaxies, however, the TF-scatter in the
current sample is rather dominated by various observational errors. Hence, its
variations can not be really used to constrain intrinsic galaxy properties like
$\Upsilon^*_K$ and $\eta$.

\begin{table}
\caption{Observational scatter in the baryonic Tully-Fisher diagram as a
function of adopted values $\Upsilon^*$ and $\eta$.}
\begin{tabular}{ccccccc}
\hline
$\eta$       & 1.33  & 1.85  & 2.50  & 3.00  & 4.00  & 5.00 \\
$\Upsilon^*$ &       &       &       &       &       &      \\
\hline
0.40         & 0.357 & 0.353 & 0.352 & 0.353 & 0.356 & 0.360\\
0.50         & 0.362 & 0.355 & 0.352 & 0.352 & 0.353 & 0.356\\
0.60         & 0.366 & 0.358 & 0.354 & 0.352 & 0.352 & 0.354\\
0.75         & 0.371 & 0.363 & 0.357 & 0.354 & 0.352 & 0.352\\
1.00         & 0.380 & 0.370 & 0.363 & 0.359 & 0.354 & 0.352\\
\hline
\end{tabular}
\end{table}

The obtained parameters of regression lines for the different kinds of TF
relations are summarized in Table~3.

\begin{table}
\caption{ The measured values of slope ($\beta$), zero-point (C) and dispersion
($\sigma$) for the various types of Tully-Fisher relations.}
\begin{tabular}{ccccccc}
\hline
LV sample                                 & N   & $\beta$ &   C   & $\sigma$ \\
\hline $M_B$ vs $W_{50}$, TRGB, $i>45$    & 206 &  -6.85  & -2.78 &   1.06   \\
$M_B$ vs $W^c_{50}$, TRGB, $i>45$         & 206 &  -6.96  & -2.19 &   1.10   \\
$L_K$ vs $W_{50}$, TRGB, $i>45$           & 206 &   2.87  &  3.16 &   0.46   \\
$L_K$ vs $W^c_{50}$, TRGB, $i>45$         & 206 &   2.91  &  2.91 &   0.47   \\
$M_{HI}$ vs $W_{50}$, TRGB, $i>45$        & 206 &   2.08  &  4.24 &   0.43   \\
$M_{HI}$ vs $W^c_{50}$, TRGB, $i>45$      & 206 &   2.14  &  4.01 &   0.44   \\
$M_{HI}$ vs $W^c_{50}$, $M_{HI}/M^*>0.04$ & 190 &   2.28  &  3.79 &   0.40   \\
$M_{bar}$ vs $W_{50}$                     & 206 &   2.55  &  3.91 &   0.36   \\
$M_{bar}$ vs $W^c_{50}$                   & 206 &   2.61  &  3.64 &   0.38   \\
$M_{bar}$ vs $W_{50}$, extended           & 331 &   2.49  &  3.97 &   0.38   \\
\hline
\end{tabular}
\end{table}

\section{A secondary parameter in the bTF-diagram.}
In literature, one can find different attempts of reducing the scatter of
galaxies in the Tully-Fisher diagram by introducing some additional parameters
(Kashibadze 2008; Kudrya et al. 2009; McCall et al. 2012). Apparently, these
parameters should be chosen as been independent on galaxy distance. The examples
of such variables are: surface brightness of a galaxy, its gas fraction index,
morphological type, apparent axial ratio, specific star formation rate, and
isolation index. Strictly speaking, all of them still show a slight dependence
on distance due to different effects of observational selection. It should be
also kept in mind that the listed parameters may be mutually correlated.
Nevertheless, the search for a secondary parameter in the bTF relation offers a
chance to improve the accuracy of measuring galaxy distances by such approach
(if the application of the corresponding parameter does not drop the sample
number essentially).

As candidates, we have tested the following parameters available in the UNGC.

 1) The mean surface brightness in $B$ band within the Holmberg isophote,
$$SB=B^c_T+5 \log a_{26}+ 8.63,$$ where the total apparent magnitude $B^c_T$ 
and the angular diameter in arcminutes, $a_{26}$, are corrected for extinction
and inclination effects.

 2) The gas fraction index ($B^c_T - m_{21}$).

 3) The specific star formation rate, $$P=\log(SFR) -\log L_K+10.14,$$ normalized 
to the age of Universe. Here, the integral star formation rate was defined either 
from apparent magnitude in far ultraviolet based on GALEX data (Gil de Paz et al.
2007) $$ \log(SFR[M_\sun\times yr^{-1}]) = 2.78 - 0.4 m^c_{FUV}+2 \log D$$ or
via $H\alpha$ flux (Karachentsev \& Kaisina 2014)

$$ \log(SFR[M_\sun\times yr^{-1}]) = 8.98 + \log F_c(H\alpha)+2 \log D $$

with considering galactic and internal extinction: $A(FUV) = 1.93 (A^G_B +
A^i_B)$ and $A(H_{\alpha}) = 0.538 (A^G_B + A^i_B)$, where $A^G_B$ from Schlafly
\& Finkbeiner (2011) and $A^i_B$ from equation (6). 

All three parameters are independent from the distance $D$, and their values 
are available in our database for most LV galaxies. 
We also tested two parameters, characterizing the local environment of a
galaxy: the index of isolation 

 $$\Theta_1= \max[\log(M_n/D^3_n)]+ C, \,\,\,\, n=1,2,...\, N, \eqno(7)$$

which distinguishes among the plenty of nearby galaxies the most significant 
neighbor (the “main disturber”=MD), whose tidal force, $F_n \sim M_n/D^3_n$ 
dominates all other neighbors (Karachentsev et al, 2013). Here, $D_n$ is 
the 3D separation of a neighboring galaxy, and the value of constant 
$C = -10.96$ was chosen so that at $\Theta_1=0$ the neighbor “n” is located on the 
“zero velocity sphere” relative to the MD. At that, the galaxies with $\Theta_1>0$
turned out to be members of a certain group, and the negative values of $\Theta_1$
corresponded to isolated galaxies. The tidal index $\Theta_1$ or actually a 
stellar density contrast, contributed by one, most important neighbor, i.e. MD,
can significantly change with time due to orbital motions of galaxies. That is
why we also used another isolation index

 $$\Theta_5=\log(\sum^5_{n=1} M_n/D_n^3)+ C,\eqno (8) $$

which is the sum of the density contrasts produced by five most important 
neighbors. The value of constant C here is the same as in Equation (7).

Four panels of Fig.~12 show the relation between a galaxy deviation from the
regression line and the value of each of four mentioned parameters. The galaxy
subsample designations are the same as in Fig.~11. Table~4 provides the values of
the regression slope $\beta$ with its error, the RMS scatter $\sigma$, and the
Pearson ($\rho$) and Kendall ($\tau$) correlation coefficients with their
statistical significance p-values. No correlation is statistically significant since
absolute values of all correlation coefficients are well below 0.2.
The observed lack of environmental dependence of the bTF relation looks to be a
rather unexpected result.

\begin{table}
\caption{Various residual non-correlations }
\begin{tabular}{lrrrrrrrr}
\hline
                 & $\beta$ & $\pm$ & $\sigma$ &  $\rho$ & $p_\rho$ & $\tau$ & $p_\tau$ \\
\hline
SB               &  -0.078 & 0.023 &   0.376  &  -0.160 &   0.003  & -0.107 &   0.004  \\
$B_T^c - m_{21}$ &  -0.034 & 0.015 &   0.380  &  -0.077 &   0.164  & -0.082 &   0.027  \\
P                &   0.006 & 0.040 &   0.380  &   0.008 &   0.891  & -0.008 &   0.825  \\
$\Theta_1$       &   0.008 & 0.014 &   0.380  &   0.019 &   0.732  &  0.038 &   0.301  \\
$\Theta_5$       &   0.008 & 0.014 &   0.380  &   0.019 &   0.742  &  0.038 &   0.301  \\
$log(b/a)$       &   0.044 & 0.102 &   0.383  &   0.018 &   0.746  & -0.029 &   0.430  \\

\hline
\end{tabular}
\end{table}

\section{Bald (dSph) dwarfs in the TF-diagram.}
As it is seen from Figures 3 \& 4, only 656 out of 1049 Local Volume objects
have been detected in HI. The fraction of objects without HI-data is growing rapidly towards the low mass
galaxies. About a third of the HI-undetected objects have been never observed in
HI because they are located outside the HIPASS and ALFALFA surveys. The majority
of the remainig undetected objects are dwarf spheroidal (dSph) and dwarf
elliptical (dE) galaxies with a small admixture of transition dwarfs (dTr) and
outlying globular clusters around M31. The relative number of true gas-poor dSph
and dE galaxies (bald dwarfs) in the LV sample amounts to $\sim20$\%. However,
this quantity is rather uncertain. No evidence of gas or structure details, as
well as low surface brightness, make dSphs very difficult to determine their
distances and radial velocities. Actually, most their distances were estimated
from supposed membership in the known groups. There is no any deep systematic
survey for dSphs over the whole sky. In practice, dSph galaxies can be
identified as the Local Volume objects only after resolving them into stars,
which is possible to reach with the existing ground-based telescopes just within
the Local Group and its immediate surroundings.

At present, radial velocities and stellar radial velocity dispersion,
$\sigma_*$, are measured for 69 dSph and 2 dE satellites of the Milky Way and
M31. We compiled a summary of their $\sigma_*$ and stellar masses, based on the
data from Collins et al. (2013, 2014) and Wheeler et al. (2015) with additions
from Walker et al.(2016) and Irwin \& Tolstoy (2002). Their distribution is
presented in Fig.~13, where horizontal bars indicate the measurement errors. As
one can see, for many tiny dwarfs velocity dispersion is comparable with its
typical error of $\sim(2-5)$~km/s.

Then, we incorporated the data on dSph and dE companions of the Milky Way and
M31 into the bTF diagram, adopting the value $W_* = 2 \sqrt{2 \ln 2}\times
\sigma_*$ as an analogue of the line width $W_{50}$ for the Gaussian function,
and setting $M_{bar}= M^*$. The combined bTF diagram for 404 late-type and
early-type galaxies is shown in Fig.~14. The gas-poor dwarfs are indicated by
red stars. As seen, the population of dSphs forms a wide tail curved downward
with much steeper bTF slope. An essential part of their scatter is caused by
errors in $\sigma_*$ themselves. In the range of $W > 20$~km/s, most the
early-type dwarfs follow nearly the same bTF relation as gas-rich galaxies, that
has been found by McGaugh \& Wolf (2010) and den Heijer et al. (2015). Although
in our opinion, the dSphs show at average lower masses than dIrs. Being improved
in $\sigma_*$ errors, this diagram will be useful to test different scenarios
describing transformation of late-type dwarfs into quenched dSphs. McGaugh \&
Wolf (2010) also noted that the deviations of dSphs from bTF relation are
considerably more susceptible to tidal effects in Modified Newtonian dynamics
(MOND) than in the standard dark matter paradigm.

The raw observational data on 402 galaxies of the Local Volume, including 71
dSphs, are compiled in Table 5. Its columns contain: (1) galaxy name; (2)
equatorial coordinates; (3) morphological type; (4) apparent axes ratio; (5) HI
line width $W_{50}$ (or $W^*$, marked by asterisk); (6) galaxy distance in Mpc
and the method used to determine it; (7) logarithm of the $K$-band luminosity in
solar units; (8) logarithm of hydrogen mass in solar masses; (9) logarithm of
baryonic mass with parameters: $\Upsilon^*= 0.60$, $\eta= 1.33$. The full text
of the Table is available online in http://www.sao.ru/lv/lvgdb/.

\section{Discussion and concluding remarks.}
Estimating the budget of errors responsible for the observed scatter of galaxies
in the bTF diagram we have taken into account five different sources of the scatter.

1) Many dwarf galaxies in the UNGC sample have eye-ball estimates of apparent
$B$ magnitude with an accuracy of $\sim0.5^m$. An additional error appears while
converting $B$ magnitudes to $K$ ones. Suggesting the total error
$\sigma(K)\simeq0.6^m$, we obtain its contribution to the baryonic mass error as
$\sigma(\log M_{bar})$ = 0.25.

2) Inspection of the HI line width errors in our sample yields the average
mean-square value of 17\% or 0.07 dex. At the slope of the regression line as
$\beta=2.6$, the contribution of $\sigma(W_{50})$ errors accounts to $\sim0.18$ dex,
i.e. slightly less than the contribution of photometric errors.

3) According to Meidt et al. (2014) the scatter on the stellar mass-to-NIR 
luminosity ratio for galaxies amounts $\sim0.11$ dex.

4) The uncertainty in distances for the LV objects $(\sim10$\%) produces the 
error in calculating baryonic mass of $\sim0.08$ dex, being lower than from
previous observables.

5) The typical observed error of HI flux for our sample galaxies is 
$\sim20\%$, that gives $\sim0.08$ dex on the bTF plot.

Therefore, after adding quadratic errors, the resulting expected error from 
these observables is $\sigma(\log M_{bar})= 0.35$. Performing its quadratic
subtraction from the observed scatter of galaxies in the bTF diagram (0.38 dex),
we obtain the intrinsic (cosmic) scatter in the baryonic mass to be $\sigma(\log
M_{bar})_{cosmic}= 0.15$. A part of this residual scatter is caused by imperfect
recipe used to account inclinations of very low mass galaxies often having
irregular shapes. 

Our evaluation of the observed (0.38 dex) and intrinsic (0.15 dex) scatter in
$M_{bar}$ for the sample of 331 nearby dwarf galaxies can be compared with other
estimates from literature. It should be kept in mind, that in most papers
devoted to baryonic Tully-Fisher relation the analysis is based on samples
consisting of spiral galaxies with typical HI line width $W_{50}>100$~km/s,
where dwarf galaxies are only a kind of supplement.

McGaugh (2012) selected 47 gas-rich galaxies in the range of $\log(M_{bar})$ = [7
- 11] $\log(M_\sun)$ and found for them the observed scatter of 0.24 dex. Just
the same observed scatter, 0.24 dex, has been obtained by McCall et al. (2012)
based on a deep $K$-band photometry of 19 late-type dwarf galaxies of the Local
Volume. Their sample has a median value of $\log(M_{bar})$ about 8.3
$\log(M_\sun)$. Lelli et al. (2016b) studied a sample of 118 spiral and irregular
galaxies with a high quality data on their photometry and extended HI rotation
curves. For this sample covering the baryonic mass range of $\log(M_{bar})$ =
[8.0 - 11.5] $\log(M_\sun)$ the authors found the observed scatter of 0.22 dex
and estimated the intrinsic scatter as 0.10 dex. Almost the same quantities of
the observed and intrinsic scatter were obtained by McGaugh \& Schombert (2015)
for a sample of 26 S and dIr galaxies with the median $V_{flat}$ velocity of
130~km/s. The sample of 97 gas-rich spiral and irregular galaxies selected by
Papastergis et al. (2016) probes the bTF relation over the $\log(M_{bar})$ = [8.5
- 10.5] $\log(M_\sun)$. The authors derived for them the observed scatter of
baryonic mass to be $\sim0.22$ dex. A close value, 0.25 dex, was found by
Bradford et al. (2015) for a sample of 148 gas-rich galaxies from SDSS with
stellar masses ranging between $10^7$ and $10^{9.5} M_\sun$. Recently, Bradford
et al. (2016) performed a comprehensive analysis of slope and dispersion of the
bTF relation using a sample of 930 isolated galaxies that have accurate
photometry from the Sloan Digital Sky Survey (SDSS). Their sample extends over
the baryonic mass $\log(M_{bar})$ = [7.4 - 11.3] $\log(M_\sun)$. The observed bTF
scatter for the total sample is found to be 0.25 dex. In the same time, the
sub-sample of 271 low-mass galaxies with $W_{20} < 200$ km/s is characterized by
larger scatter of 0.41 dex.

We suppose that improving the measurement accuracy of $K$ magnitudes as well as
$W_{50}$ widths, expected in forthcoming deep optical, NIR, and HI sky surveys,
may reduce the observed scatter of $\log(M_{bar})$ for the low-mass dwarfs till
0.20 dex, and thereby achieve the accuracy of bTF distances for them $\sim0.10$
dex, i.e. $\sim25$\%. This improvement will be important to refine the peculiar
velocity field in the Local Universe.

		Acknowledgements.

We thank the anonymous referee fo numerous helpful comments that essentially
improved the paper. We thank Jayaram Chengalur, Brent Tully, Valentina
Karachentseva and Dmitry Makarov for fruitful discussions. This work was 
supported by the Russian Science Foundation, grant 14--12--00965. Support for
proposals GO 12546, 14636 was provided by NASA through grants from the
Space Telescope Science Institute, which is operated by the Association
of Universities for Research in Astronomy, Inc., under NASA contract NAS5–26555.
Observations with the Russian 6-meter telescope BTA and Indian Giant Meterwave
Radio Telescope GMRT have been supported by the Russian Foundation for Basic 
Research grant 15--52--45004 (IND--RUS).

		 		 R e f e r e n c e s

Aaronson M., Huchra J., Mould J. 1979, {\em ApJ}, 229, 1

Aaronson M., Bothun G., Mould J. et al. 1986, {\em ApJ}, 302, 536

Balkowski C., Bottinelli L., Chamaraux P., et al, 1974, {\em A \& A}, 34, 43

Begum A., Chengalur J.N., Karachentsev I.D., Sharina M.E., Kaisin S.S. 2008a,
{\em MNRAS}, 386, 1667

Begum A., Chengalur J.N., Karachentsev I.D., Sharina M.E. 2008b, {\em MNRAS},
386, 138

Bell E.F., McIntosh D.H., Katz N., Weinberg M.D. 2003, {\em ApJS}, 149, 289

Bradford J.D., Geha M.C., Blanton M.R. 2015, {\em ApJ}, 809, 146

Bradford J.D., Geha M.C., van den Bosch F.C., 2016, 1602.02757

Brook C.B., Santos-Santos I., Stinson G., 2016, {\em MNRAS}, 459, 638

Brook C.B., Shankar F., 2016, {\em MNRAS}, 455, 3841

Catinella B., Kauffmann G., Schiminovich D., et al. 2012, {\em MNRAS}, 420, 1959

Collins M.L., Chapman S.C., Rich R.M., et al. 2013, {\em ApJ}, 768, 172

Collins M.L., Chapman S.C., Rich R.M., et al. 2014, {\em ApJ}, 783, 7

de Vaucouleurs, G., de Vaucouleurs, A., \& Corwin, H. 1976, Second Reference
Catalogue of Bright Galaxies, Texas, Austin

den Heijer M., Oosterloo T.A., Serra P., et al. 2015, {\em A\&A}, 581A, 98

Fukugita M., Peebles P.J.E. 2004, {\em ApJ}, 616, 643

Garcia-Ruiz \& Sancisi 2002, {\em A\&A}, 394, 769

Geha M., Blanton M.R., Masjedi M., West A.A. 2006, {\em ApJ}, 653, 240

Gil de Paz, A., Boissier, S., Madore, B. F., et al. 2007, {\em ApJS}, 173, 185

Giovanelli R., Haynes M.P., Herter T., et al. 1997, {\em AJ}, 113, 53

Giovanelli R., Haynes M.P., Kent B.R. et al. 2005, {\em AJ}, 130, 2598

Gurovich S., Freeman K., Jerjen H., et al. 2010, {\em AJ}, 140, 663

Haynes M.P., Giovanelli R., Martin A.M. et al. 2011,{\em AJ}, 142, 170

Huchtmeier, W.K., Karachentsev, I.D., Karachentseva, V.E. \& Ehle M. 2000, {\em
A\&AS}, 141, 469

Huchtmeier W.K., Karachentsev I.D., Karachentseva V.E. 2001, {\em A\&A}, 377,
801

Huchtmeier W.K., Karachentsev I.D., Karachentseva V.E. 2003, {\em A\&A}, 401,
483

Irwin M., Tolstoy E. 2002, {\em MNRAS}, 336, 643

Jarrett, T.N., Chester, T., Cutri R. et al. 2000, {\em AJ}, 119, 2498

Jarrett T.H., Chester T., Cutri R., Schneider S. E., Huchra J. P. 2003, {\em
AJ}, 125, 525

Just A., Fuchs B., Jahreiss H., et al. 2015, {\em MNRAS}, 451, 149

Kaisina E.I., Makarov D.I., Karachentsev, I.D., Kaisin S.S. 2012, {\em AstBu},
67, 115

Karachentsev I.D., 1989, {\em AJ}, 97, 1566

Karachentsev I.D., Karachentseva V.E., Kudrya Yu.N., 1999, Astronomy Letters,
25, 1

Karachentsev I.D., Mitronova S.N., Karachentseva V.E., Kudrya Yu.N., Jarrett
T.H. 2002, {\em A\&A}, 396, 431

Karachentsev, I.D., Karachentseva, V.E., Huchtmeier, W.K., Makarov, D.I. 2004,
{\em AJ}, 127, 2031 (CNG)

Karachentsev I.D., Makarov D.I., Kaisina E.I. 2013, {\em AJ}, 145, 101 (UNGC)

Karachentsev, I.D., Kaisina, E.I. 2014, {\em AJ}, 146, 46

Karachentseva V.E., Kudrya Y.N., Karachentsev I.D., et al. 2016, {\em AstBull},
71, 1

Kashibadze O.G., 2008, Ap, 51, 336

Kirby E.M., Koribalski B., Jerjen H., Sanchez A.L. 2012, {\em MNRAS}, 420, 2924

Koribalski B.S., Staveley-Smith L., Kilborn V.A. et al. 2004, {\em AJ}, 128, 16

Kova\^{c} K., Oosterloo T.A., van der Hulst J.M. 2009, {\em MNRAS}, 400, 743

Kraan-Korteweg, R.C. \& Tammann, G.A. 1979, {\em AN}, 300, 181

Kudrya Yu. N. Karachentseva V.E., Karachentsev I.D., et al. 2009, {\em Ap}, 52,
367

Lelli F., Verheijen M., Fraternali F., 2014, A \& A, 566A, 71

Lelli F., McGaugh S.S., Schombert J.M., Pawlowski M.S., 2016a, {\em Ap}, 827L,
19

Lelli F., McGaugh S.S., Schombert J.M. 2016b, {\em ApJ}, 816, L14

Martin N.F., de Jong J.T.A., Rix H.W., 2008, {\em ApJ}, 684, 1075

McCall M.L., Vaduvescu O., Nunez F.P., et al. 2012, {\em A\&A}, 540A, 49

McGaugh S.S. 2005, {\em ApJ}, 632, 859

McGaugh S.S. 2012, {\em AJ}, 143, 40

McGaugh S.S., Schombert J.M., Bothun G.D., de Blok W.J. 2000, {\em ApJ}, 533,
L99

McGaugh S.S., Schombert J.M., de Blok W.J., Zagursky J. 2010, {\em ApJ}, 708L,
14

McGaugh, S.S., \& de Blok, W.J.G. 1997, {\em ApJ}, 481, 689

McGaugh S.S., Schombert J.M. 2014, {\em AJ}, 148, 77

McGaugh S.S., Schombert J.M. 2015, {\em ApJ}, 802, 18

McGaugh S.S., Wolf J., 2010, {\em ApJ}, 722, 248

Meidt S.E., Schinnerer E., van de Ven G., et al. 2014, {\em ApJ}, 788, 144

Meyer M.J., Zwaan M.A., Webster R.L. et al. 2004, {\em MNRAS}, 350, 1195

Miller M.J., Bregman J.N. 2015, {\em ApJ}, 800, 14

Noordermeer E., Verheijen M.A.W., 2007, {\em MNRAS}, 381, 1463

Obreschkow D., Meyer M. 2013, {\em ApJ}, 777, 140

Papastergis E., Adams E.A.K., van der Hulst J.M. 2016, a-ph/1602.09087

Paturel G., Andernach H., Bottinelli L., et al. 1997, {\em A\&AS}, 124, 109

Pfenniger D., Revaz Y. 2005, {\em A\&A}, 431, 511

Reyes R., Mandelbaum R., Gunn J.E., et al. 2011, {\em MNRAS}, 417, 2347

Roberts M.S. 1969, {\em AJ}, 74, 859

Roychowdhury S., Chengalur J.N., Begum A., Karachentsev I.D. 2010, {\em MNRAS},
404, L60

Roychowdhury S., Chengalur J.N., Karachentsev I.D., Kaisina E.I. 2013, {\em
MNRAS}, 436L, 104

Sales V.L., Navarro J.F., Oman K. et al., 2016, a-ph/1602.02155

Sanchez-Janssen R., Mendez-Abreu J., Aguerri J.A.L., 2010, {\em MNRAS}, 406L, 65

Sanchez-Janssen R., Ferrarese L., MacArthur L.A., et al, 2016, {\em ApJ}, 820,
69

Schlafly, E. F., \& Finkbeiner, D. P. 2011, {\em ApJ}, 737, 103

Schombert J., 2011, a-ph/1107.1728

Stark D.V., McGaugh S.S., Swaters R.A. 2009, {\em AJ}, 138, 392

Staveley-Smith L., Kraan-Korteweg R.C., Schroder A.C., et al. 2016, {\em AJ},
151, 52

Stocke J.T., Keeney B.A., Danforth C.W., et al. 2013, {\em ApJ}, 763, 148

Swaters R.A., Sancisi R., van Albada T.S., van der Hulst J.M., 2009, A\& A, 493,
871

Tonry, J.L., Dressler, A., Blakeslee, J.P. et al. 2001, {\em ApJ}, 546, 681

Torres-Flores S., Epinat B., Amram P. et al. 2011, {\em MNRAS}, 416, 1936

Tully R.B., Fisher J.R. 1977, {\em A\&A}, 54, 661

Tully R.B., Pierce M.J. 2000, {\em ApJ}, 533, 744

Verheijen M.A.W. 2001, {\em ApJ}, 563, 694

Verheijen M.A.W., Sancisi R., 2001, A \& A, 370, 765

Walker M.G., Mateo M., Olszewski E.W., et al. 2016, {\em ApJ}, 819, 53

Wheeler C., Pace A.B., Bullock J.S. et al. 2015, a-ph/1511.01095

Wong O.I., Ryan-Weber E.V., Garcia-Appadoo et al. 2006, {\em MNRAS}, 371, 1855

Young T., Jerjen H., Lopez-Sanchez A.R., Koribalski B.S. 2014, {\em MNRAS}, 444,
3052

Young J.S., Knezek P.M. 1989, {\em ApJ}, 347L, 55

Yuan Q.R., Zhu C.X., 2004, Chinese Astr. Astrophys., 28, 127

Yun M.S., Ho P.T.P., Lo K.Y., 1994, Nature, 372, 6506, 530

Zasov A.V. 1974, {\em AZh}, 51, 1225

\clearpage
\begin{figure} 
\epsscale{1.0}
\plotone{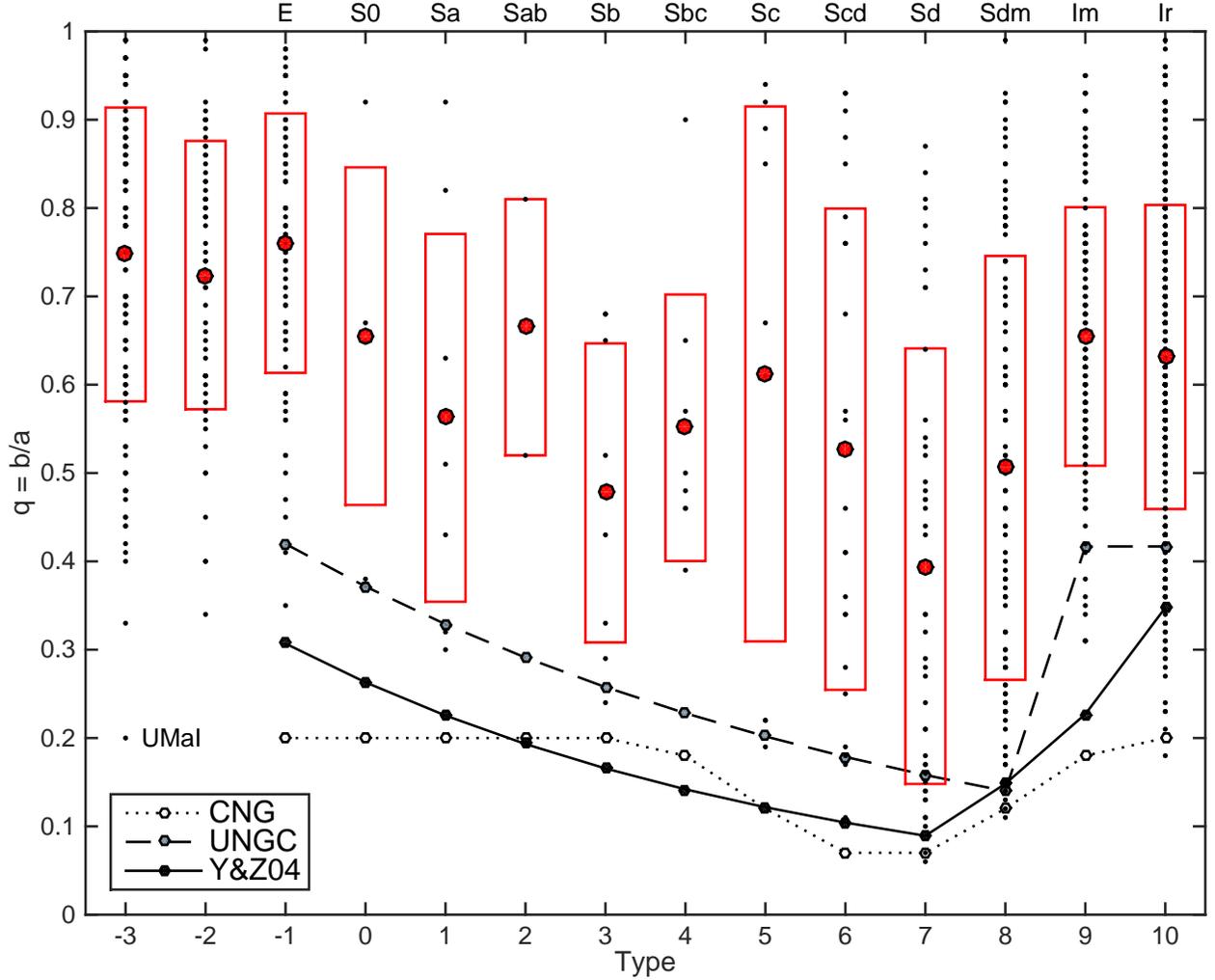}
\caption{Apparent axial ratio of the Local Volume galaxies vs. their
morphological type in de Vaucouleurs scale. The large circles and vertical
rectangles indicate the mean values and the standard deviations for each type.
The dotted, dashed and solid lines present intrinsic axial ratios for different
morphological types as adopted in the Catalog of Neighboring Galaxies, the Updated
Nearby Galaxy Catalog and as assumed by Yuan \& Zhu (2004), respectively.}
\end{figure}

\begin{figure} 
\epsscale{1.0}
\plotone{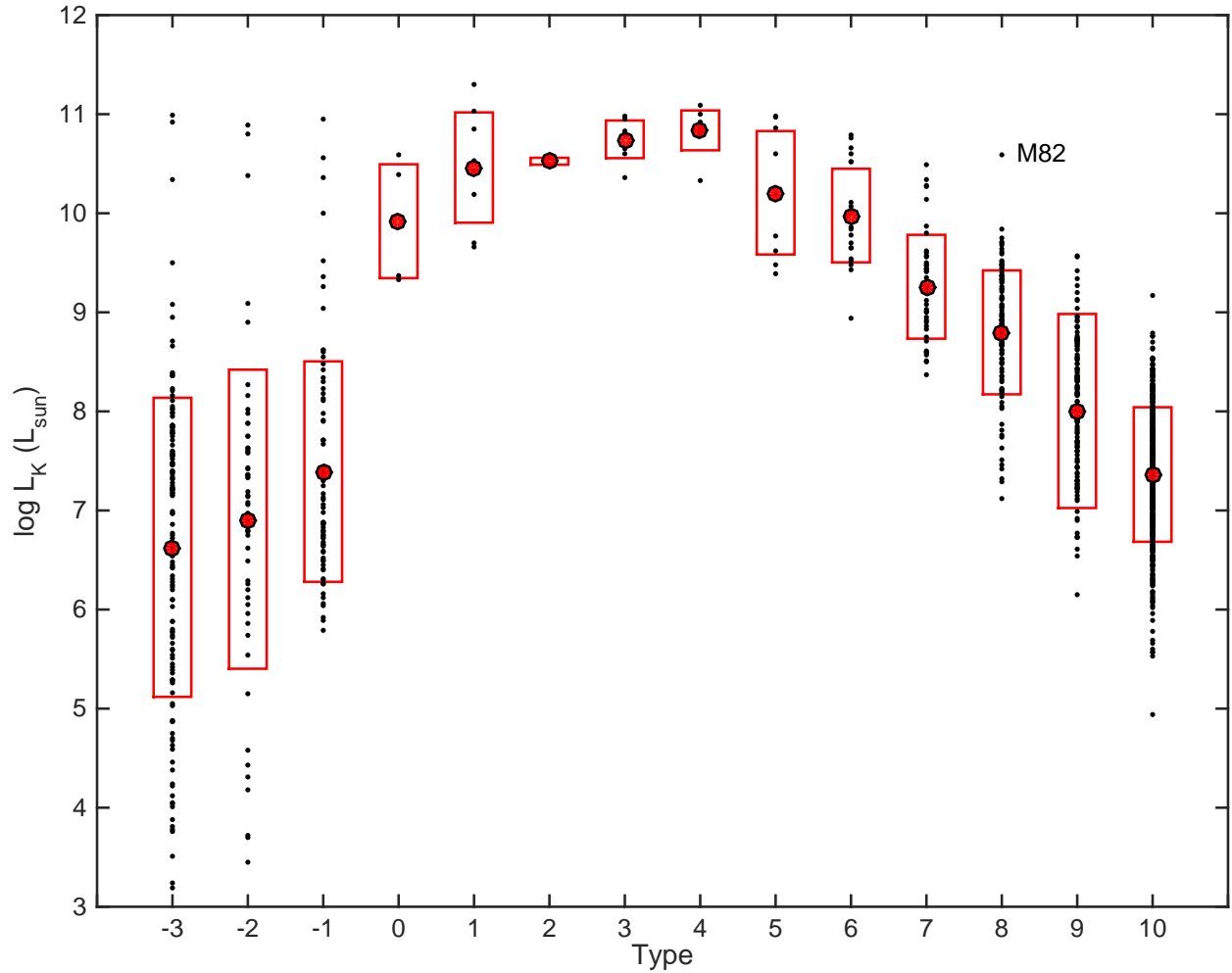}
\caption{Distribution of the LV galaxies on K-band luminosity and morphological
type. The mean values of $\log L_K$ and standard deviations are marked by large
circles and rectangles.}
\end{figure}

\begin{figure} 
\epsscale{1.0}
\plotone{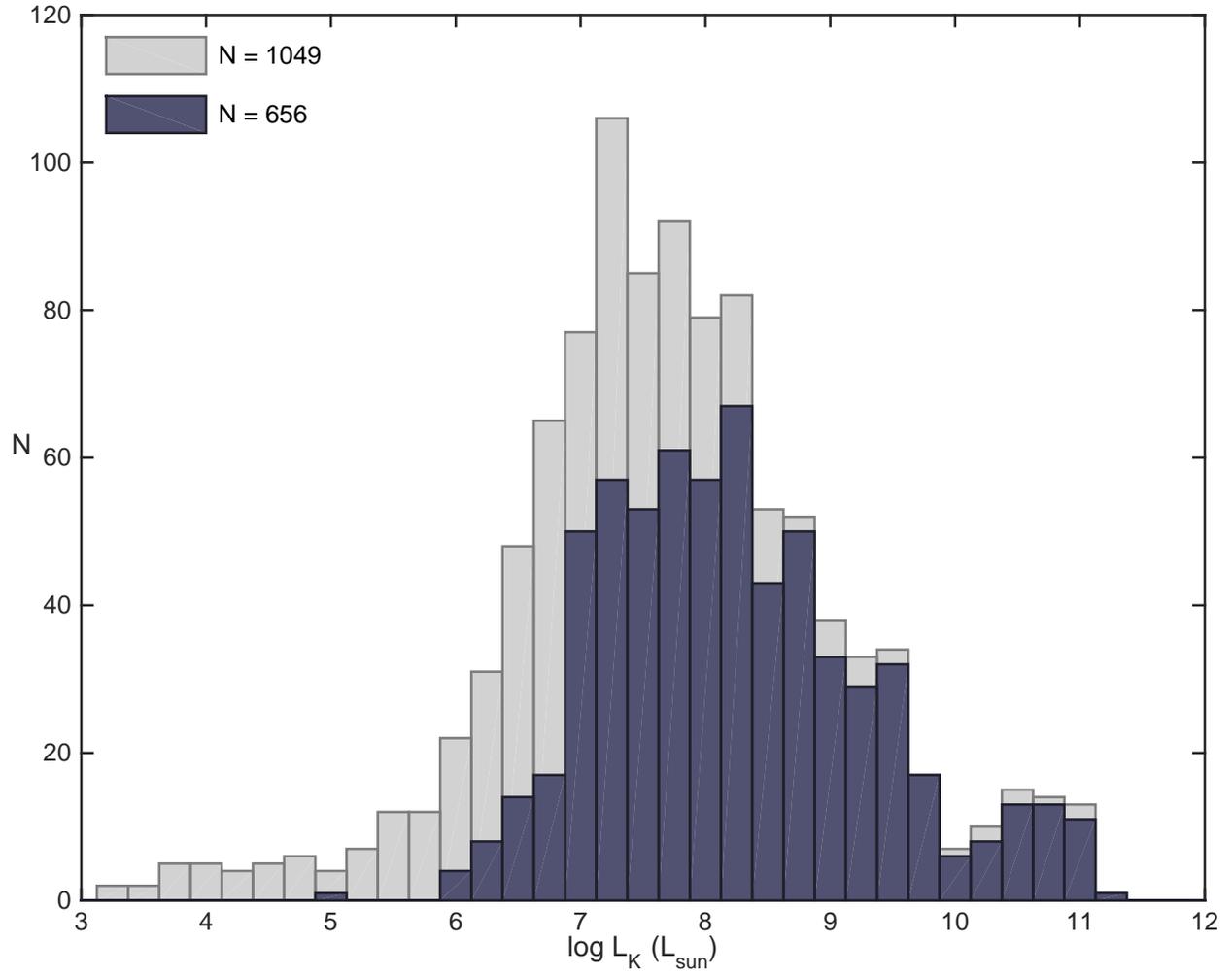}
\caption{Number of the LV galaxies vs. their K-band luminosity. HI-detected
galaxies are marked by dark color.}
\end{figure}

\begin{figure} 
\epsscale{1.0}
\plotone{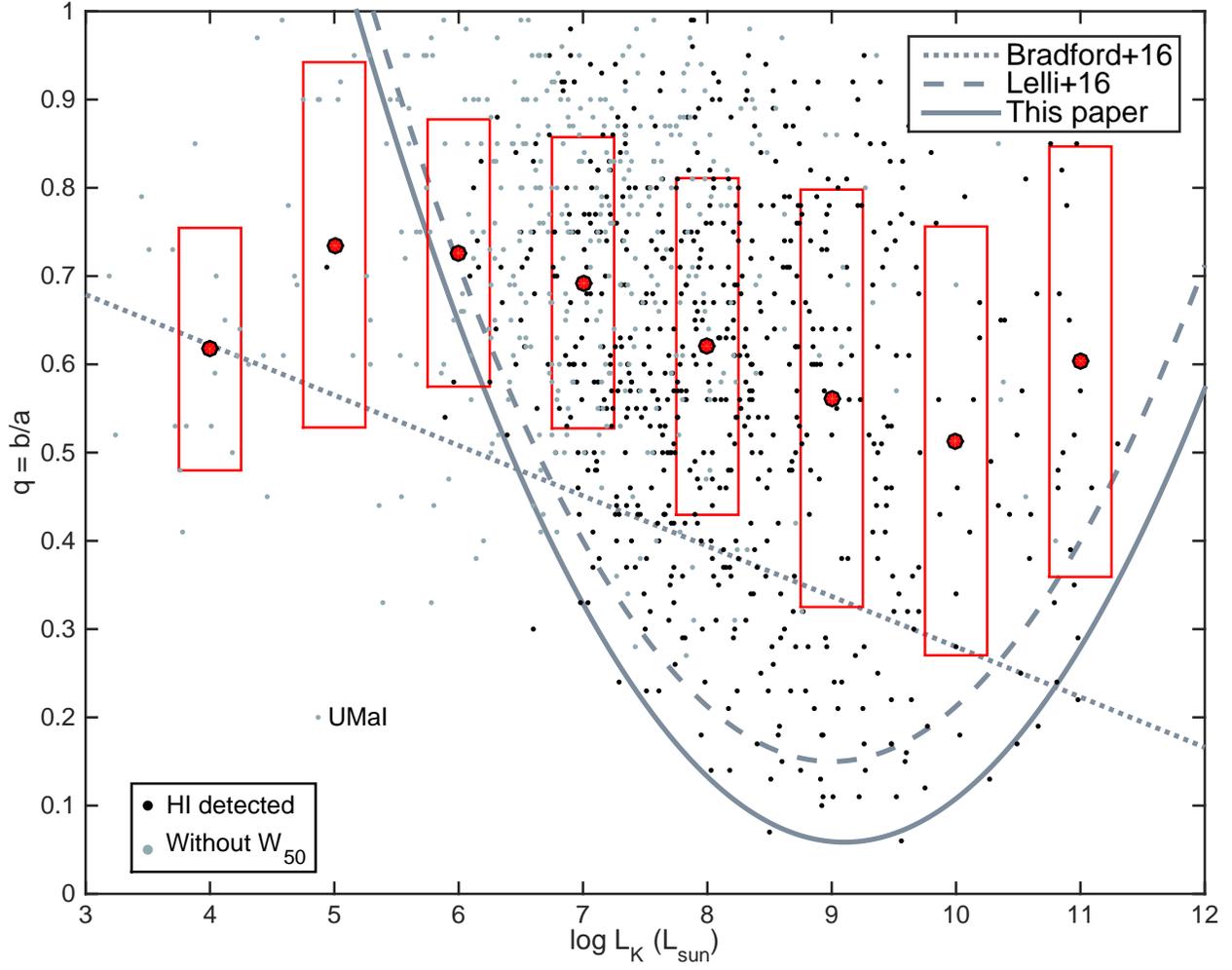}
\caption{Distribution of the LV galaxies on their apparent axial ratio and
K-band luminosity. Light and dark small circles indicate HI-detected and
non-detected galaxies respectively. The dotted, dashed and solid lines present 
intrinsic axial ratios under different assumptions.}
\end{figure}

\begin{figure} 
\epsscale{1.0}
\plotone{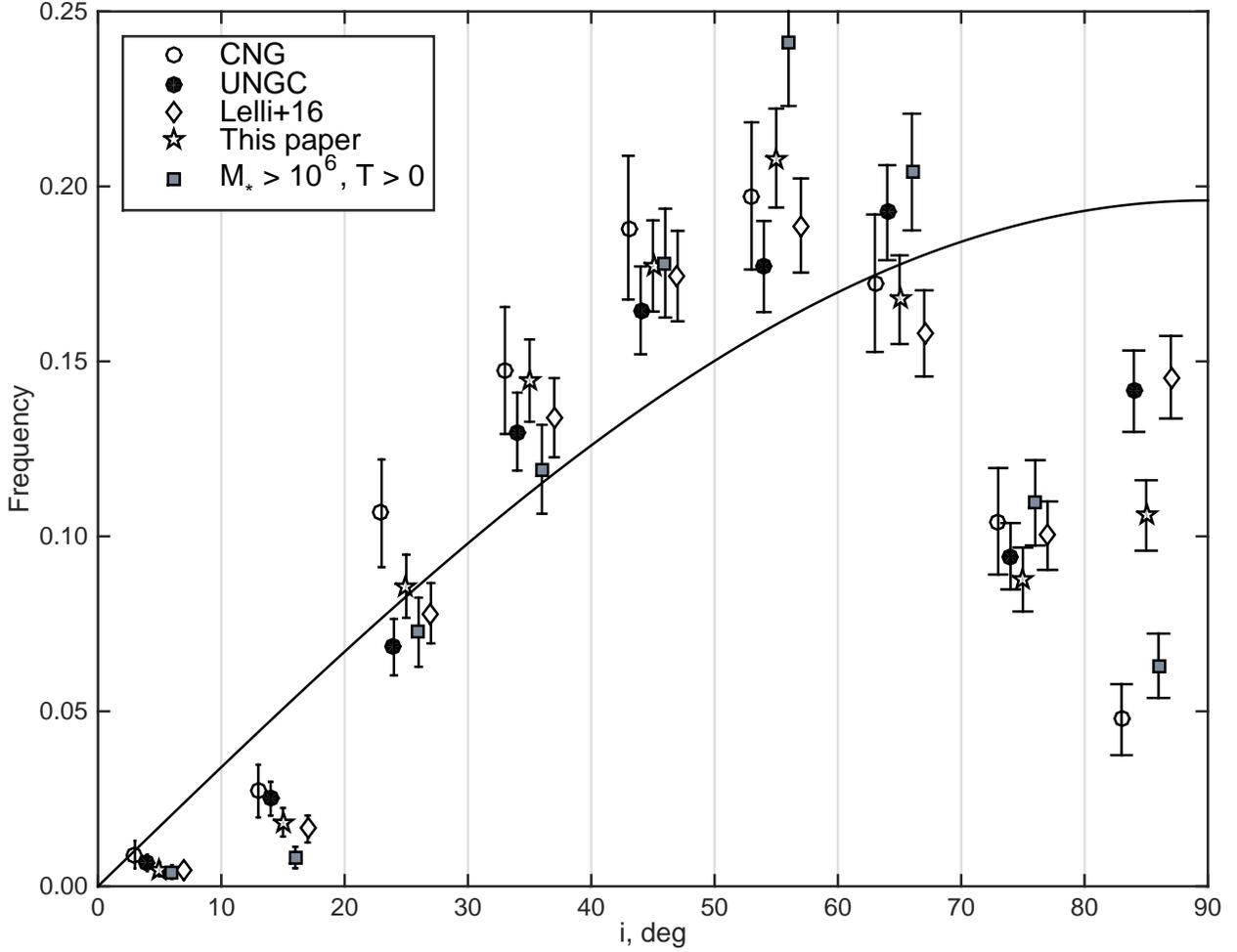} 
\caption{The frequency distribution of the LV galaxies on inclination angle $i$
(degree) under different assumptions (2)-(5) about intrinsic axial ratios. The
solid line correspondes to the $\sin(i)$ --- law expected for arbitrary
orientation of galaxy axis.}
\end{figure}

\begin{figure}
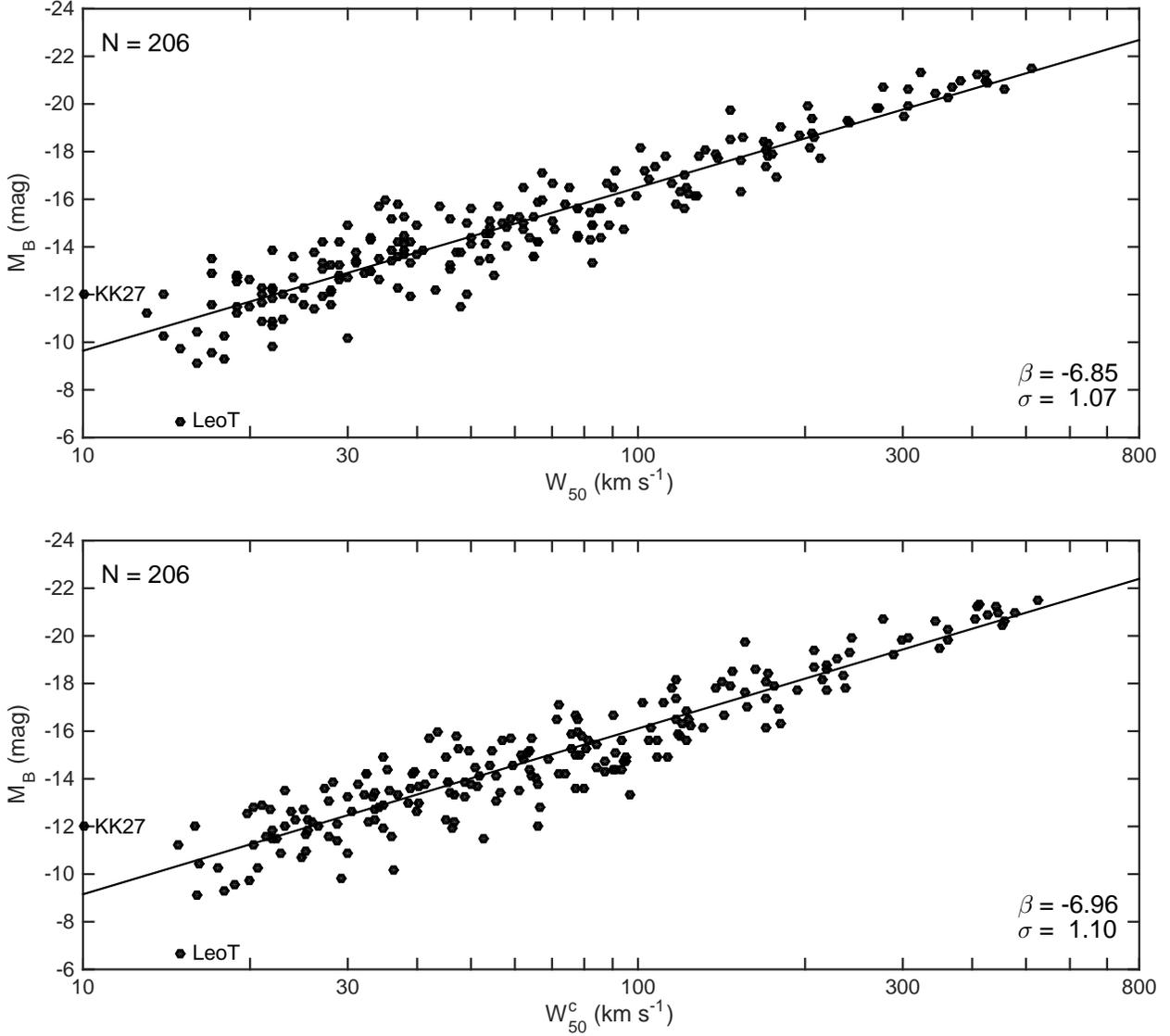
 
\epsscale{1.0}
\plotone{Fig6-1.eps}\\
\vspace{0.4cm}
\plotone{Fig6-2.eps}
\caption{$B$ absolute magnitude - line width relation for 206 LV galaxies with
accurately measured distances mainly from application of the TRGB and
inclinations $i > 45^{\circ}$. The absolute magnitudes are corrected for
Galactic and internal extinction. Horizontal scales indicate: observed line
width $W_{50}$ (upper panel), line width corrected for inclination (bottom
panel). The stright lines are least-squares fits to the ensemble with errors in
magnitudes, having a slope $\beta$ and dispersion $\sigma$.}
\end{figure}

\begin{figure}
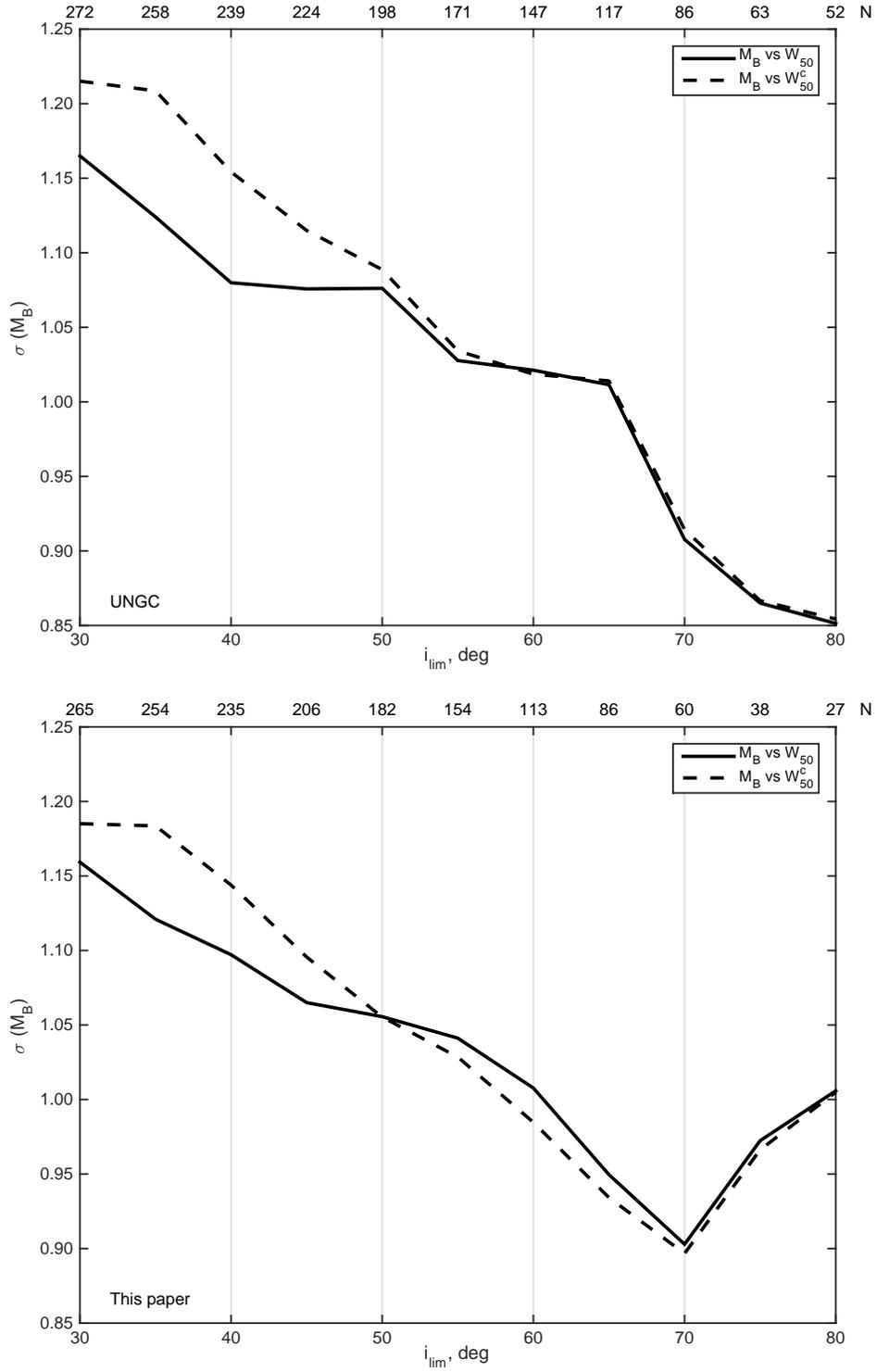
 
\epsscale{0.75}
\plotone{Fig7-1.eps}\\
\vspace{0.4cm}
\plotone{Fig7-2.eps}
\caption{ Scatter on the TF diagram as a function of limiting inclination angle
$i_{lim}$. The top and bottom panels correspond to the equations (2), and (5),
respectively.}
\end{figure}

\begin{figure}
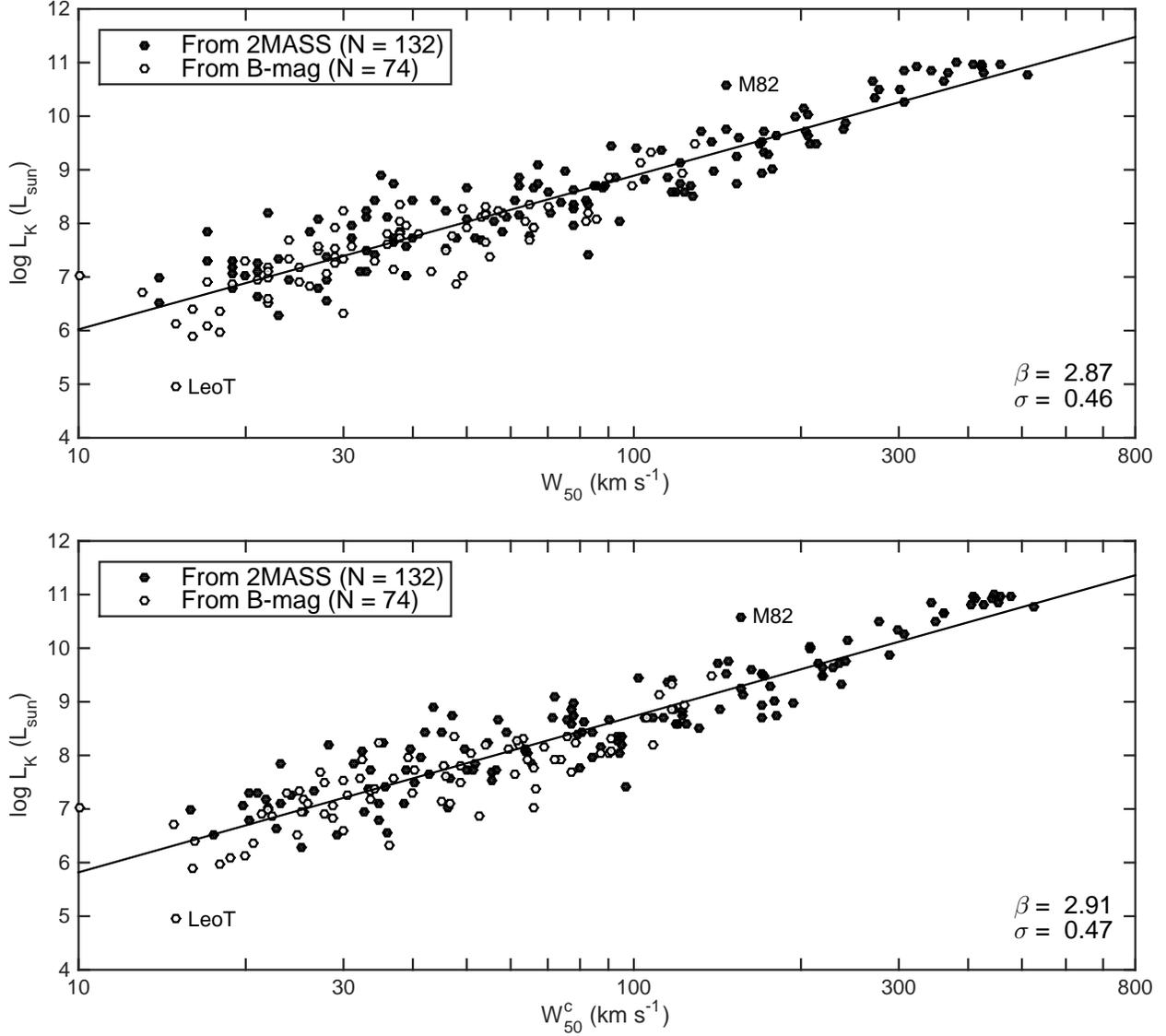
 
\epsscale{1.0}
\plotone{Fig8-1.eps}\\
\vspace{0.4cm}
\plotone{Fig8-2.eps}
\caption{$K$-band luminosity versus line width for 206 LV galaxies with
accurately determined distances and $i > 45^{\circ}$. Upper panel indicates
observed line width, bottom panel displays the line width corrected for
inclination. Filled circles correspond to 132 galaxies with $K_s$ magnitudes
from 2MASS and open circles mark 74 galaxies whose $K$ magnitudes are
estimated via the mean $B-K$ color and morphological type. Parameters $\beta$
and $\sigma$ in corners indicate the regression slope and dispersion.}
\end{figure}

\begin{figure}
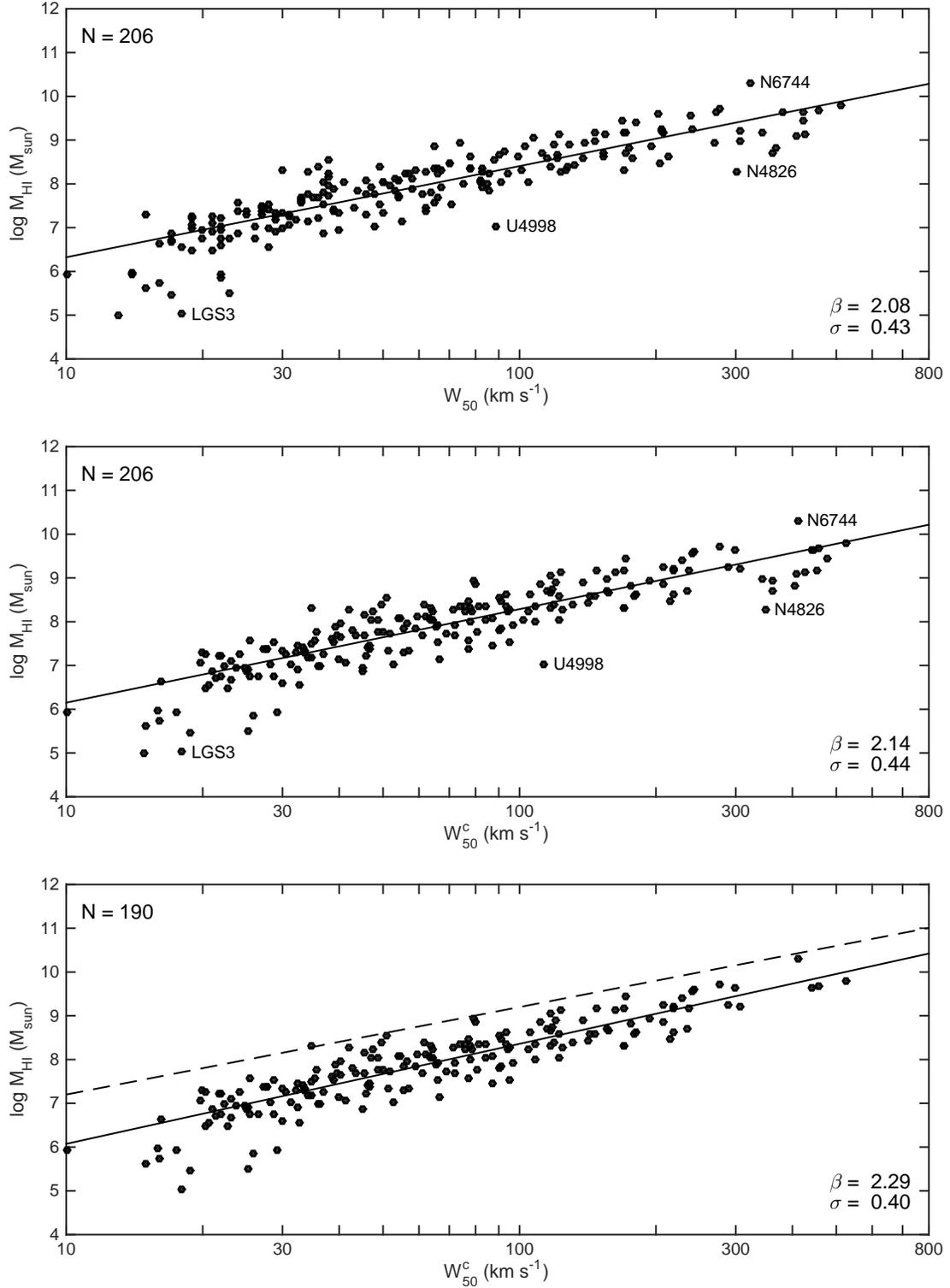
 
\epsscale{0.88}
\plotone{Fig9-1.eps}\\
\vspace{0.4cm}
\plotone{Fig9-2.eps}\\
\vspace{0.4cm}
\plotone{Fig9-3.eps}
\caption{The relation between hydrogen mass and line width for LV galaxies with
accurately measured distances. Upper and middle panels indicate line width
observed and corrected for inclination, respectively. Bottom panel corresponds
to 190 gas-rich galaxies with $m_{21} - K < 6.0^m$. The dashed line indicates
the upper limit for the ratio $M_{HI}/W_{50}^2$ interpreted as a threshould of
gravitation instability favouring star formation.}
\end{figure}

\begin{figure}
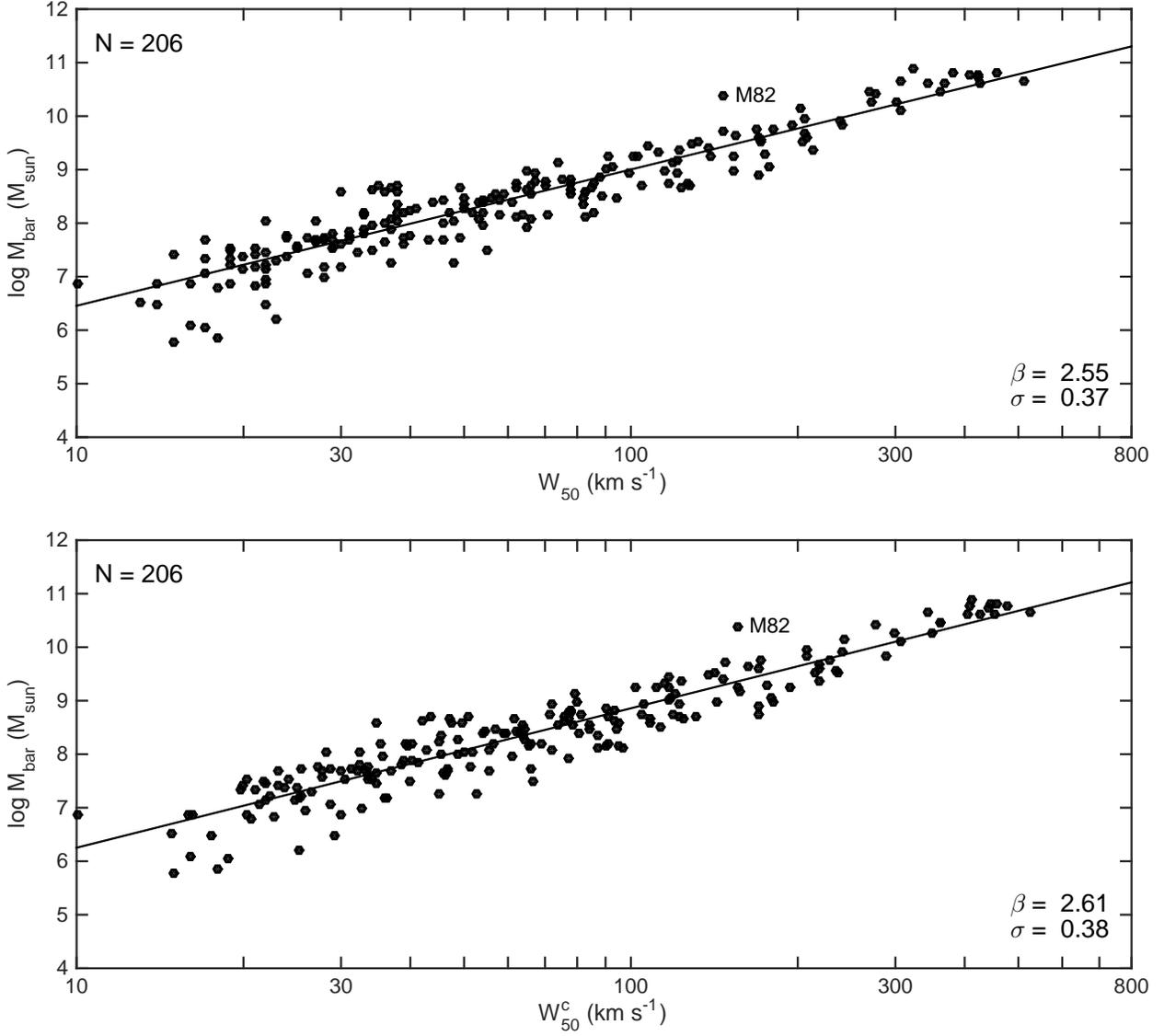
 
\epsscale{1.0}
\plotone{Fig10-1.eps}\\
\vspace{0.4cm}
\plotone{Fig10-2.eps}
\caption{Baryonic mass versus observed (upper panel) and corrected for
inclination (bottom panel) line width for 206 LV galaxies with accurate
distances. The adopted value of stellar mass-to-luminosity ratio is $\Upsilon^*
= 0.60$ and gaseous mass-to-HI-mass ratio $\eta = 1.33$.}
\end{figure}

\begin{figure} 
\epsscale{1.0}
\plotone{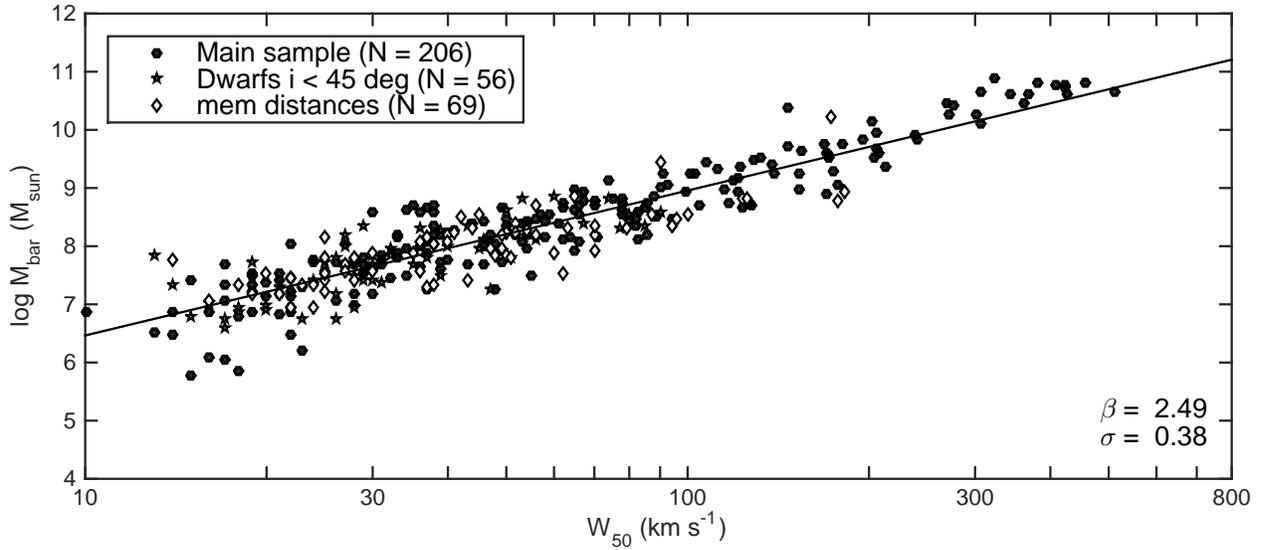}
\caption{The baryonic TF relation for 206 LV galaxies with accurately measured
distances (filled circles), 56 low luminosity dwarfs with $ i < 45^{\circ}$
(crosses) and 69 dwarfs with distances determined via membership in the known
groups (diamonds).}
\end{figure}

\begin{figure}
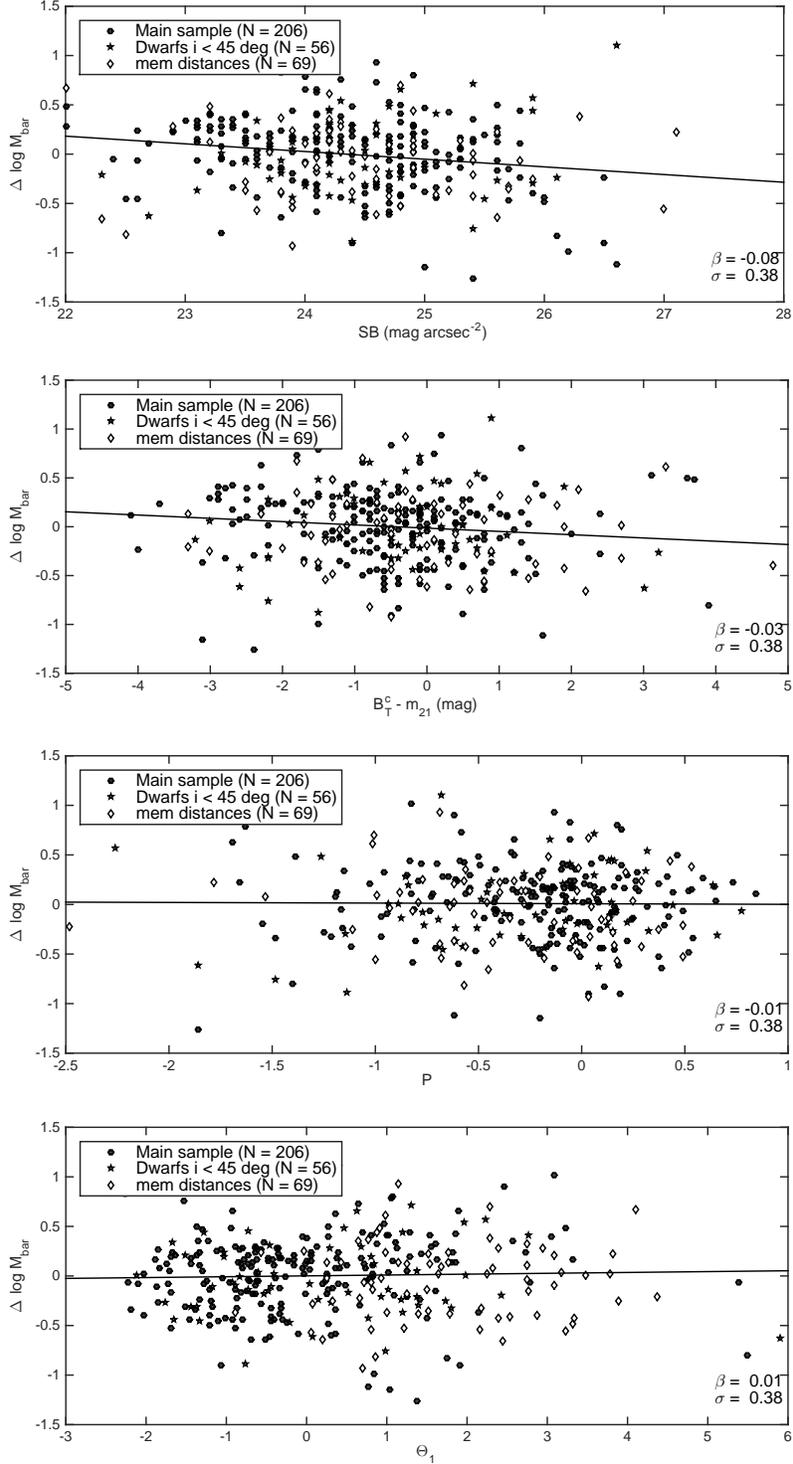
 
\epsscale{0.63}
\plotone{Fig12-1.eps}\\
\vspace{0.4cm}
\plotone{Fig12-2.eps}\\
\vspace{0.4cm}
\plotone{Fig12-3.eps}\\
\vspace{0.4cm}
\plotone{Fig12-4.eps}
\caption{Residuals in the baryonic TF diagram as a function of the galaxy
surface brightness (upper panel), ($B-m_{21}$) gas fraction index (upper middle panel),
specific star formation rate expressed in the age of Universe (lower middle panel)
and isolation index (bottom panel).
Designation of galaxy sub-samples is the same as in Fig.11. The stright lines
display regression lines with corresponding slope $\beta$ and scatter $\sigma$.}
\end{figure}

\begin{figure} 
\epsscale{1.0}
\plotone{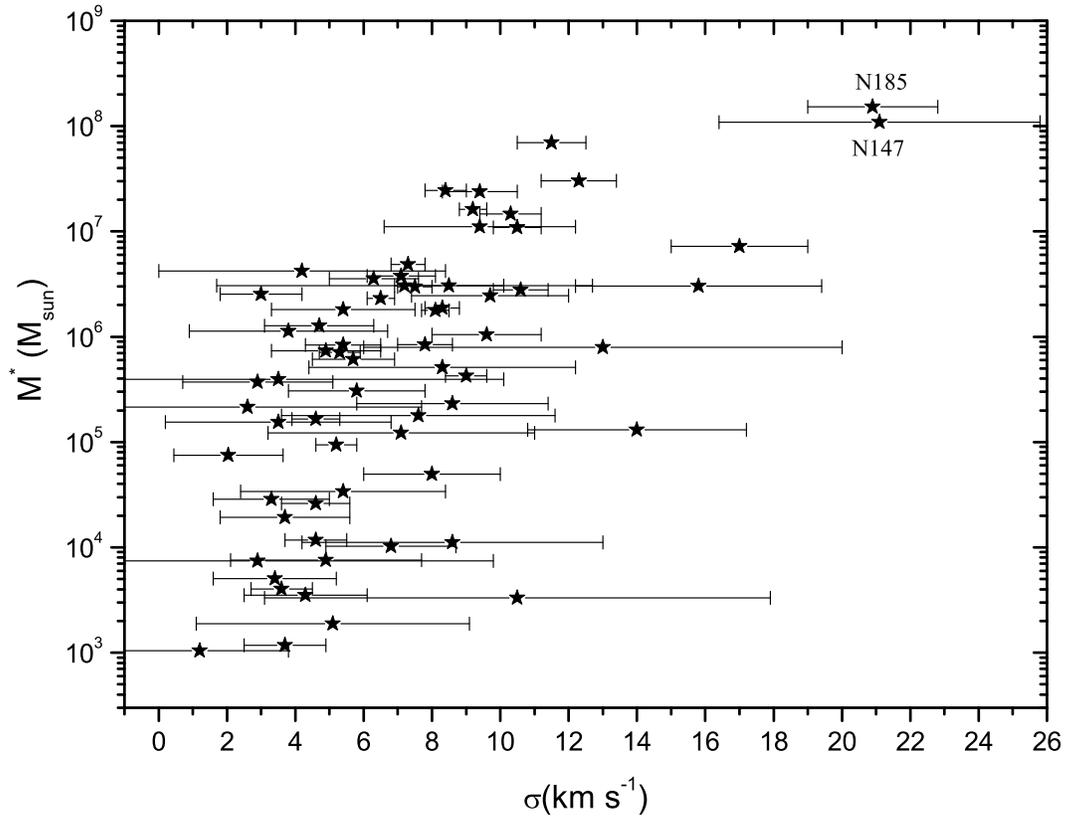}
\caption{Stellar mass versus stellar radial velocity dispersion for 69 dSph and
2 dE galaxies around the Milky Way and M31. Horisontal bars indicate measurement
errors.}
\end{figure}

\begin{figure} 
\epsscale{1.0}
\plotone{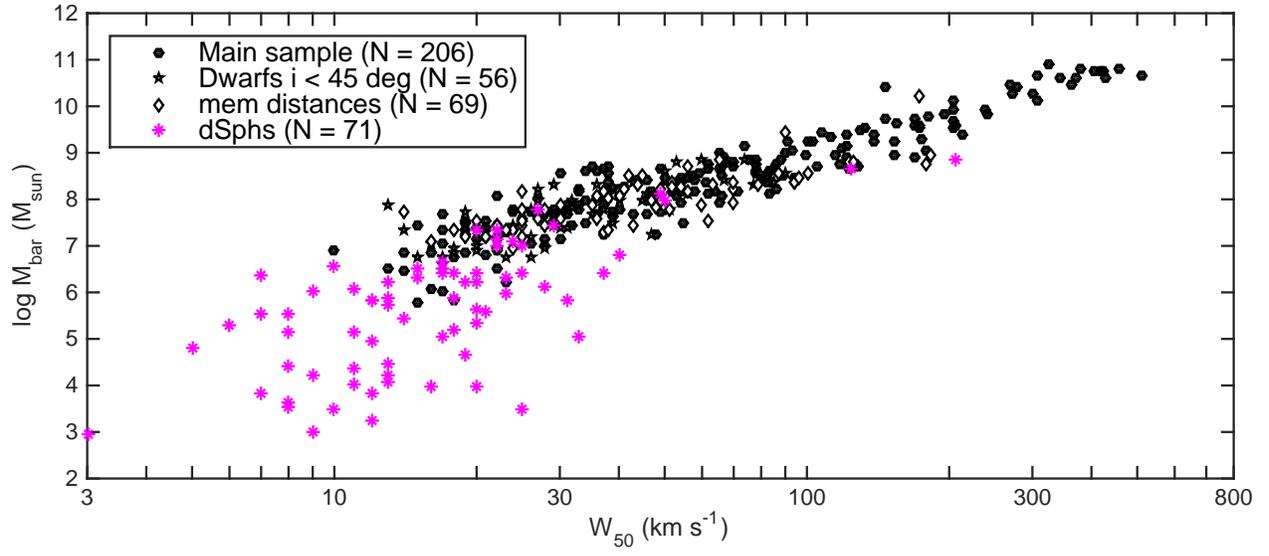}
\caption{The baryonic TF relation for 402 LV galaxies. The dSph companions of
the Milky Way and M31 are indicated by red stars.}
\end{figure}

\clearpage
\scriptsize
\begin{table}[hbt]
\caption{The initial data for the Local Volume galaxies. Table 5 is published 
      in its entirety in the machine-readable format.
      A portion is shown here for guidance regarding its form and content}
\begin{tabular}{lcrclllrr}\hline

 name & j2000 & T & b/a & $W_{50}$ & D method & $\log(L_K)$ & $\log M_{HI}$ & $M_{bar}$\\
\hline
(1)&(2)&(3)&(4)&(5)&(6)&(7)&(8)&(9)\\ \hline

WLM & 000158.1-152740 & 9 & 0.35 & 53 & 0.98 TRGB & 7.70 & 7.84 & 8.09 \\
And XVIII & 000214.5+450520 & --3 & 0.99 & 23* & 1.31 TRGB & 6.56 & 6.65 & 6.34 \\
ESO409-015 & 000531.8-280553 & 9 & 0.46 & 53 & 8.71 TRGB & 8.10 & 8.10 & 8.39 \\
AGC748778 & 000634.4+153039 & 10 & 0.52 & 16 & 6.22 TRGB & 6.39 & 6.64 & 6.86 \\
And XX & 000730.7+350756 & --3 & 0.70 & 17* & 0.80 TRGB & 5.26 & & 5.04 \\
\hline
 \end{tabular}
 \end{table}

\end{document}